\documentclass[apj]{emulateapj}
\usepackage{epsfig,amssymb}
\def\l{{\ell}}

\def\erf{\mathrm{erf}}

\def\vv{{\mathbf v}}

\def\nod{...}

\def\e{\mathrm{e}}
\def\ud{\mathrm{d}}
\def\ltsima{$\; \buildrel < \over \sim \;$}
\def\lsim{\lower.5ex\hbox{\ltsima}}

\shorttitle{Star count profiles and structural parameters of 26 GCs}
\shortauthors{P. Miocchi et al.}

\begin{document}
\defcitealias{mclaugh}{MvM05}
%\defcitealias{mio}{PM06}

%\date{12 Jan 201328 dec oct 2012 Accepted ????. Received ????; in original form ????}

%\label{firstpage}
\title
%{A structural parameters catalog for 26 Galactic globular clusters derived from complete
%star density profiles}
{Star count density profiles and structural parameters \\of 26 Galactic globular clusters}

%???? CORREGGI INDIRIZZI

\author{P. Miocchi$^1$, B. Lanzoni$^1$, F.R. Ferraro$^1$, E. Dalessandro$^1$, E. Vesperini$^2$, M. Pasquato$^3$, G. Beccari$^4$, C. Pallanca$^1$, and N. Sanna$^1$}
\email{paolo.miocchi@unibo.it}

\affil{$^1$Dipartimento di Fisica e Astronomia, Universit\`a di Bologna, Viale Berti Pichat 6/2, 40127 Bologna, Italy\\
$^2$Department of Astronomy, Indiana University, Bloomington, Indiana 47405, USA\\
$^3$Department of Astronomy \& Center for Galaxy Evolution Research, Yonsei University,
and Yonsei University Observatory,\\ Seoul 120-749, Republic of Korea\\
%\affil{Yonsei University Observatory, Seoul 120-749, Republic of Korea}
$^4$European Organization for Astronomical Research in the Southern Hemisphere,\\ K. Schwarzschild-Str. 2, 85748 Garching, Germany}

\begin{abstract}
We used a proper combination of high-resolution HST observations and
wide-field ground based data to derive the radial star density profile
of 26 Galactic globular clusters from resolved star counts
(which can be all freely downloaded on-line).
With respect to surface brightness (SB) profiles (which can be biased by the
presence of sparse, bright stars), star counts are considered to be
the most robust and reliable tool to derive cluster structural parameters.
For each system a detailed comparison with both King and Wilson models
has been performed and the most relevant best-fit parameters have been
obtained. This is the largest homogeneous catalog collected so far of
star count profiles and structural parameters derived therefrom.
The analysis of the data of our catalog has shown that: 
(1) the presence of the central cusps previously detected in the
SB profiles of NGC 1851, M13 and M62 is not confirmed; (2) the majority
of clusters in our sample are fitted equally well by the King and the
Wilson models;
(3) we confirm the known relationship between cluster size (as measured
by the effective radius) and galactocentric distances;
(4) the ratio between the core and
the effective radii shows a bimodal distribution, with a peak at $\sim
0.3$ for about $80$\% of the clusters, and a secondary peak at $\sim
0.6$ for the remaining $20$\%. Interestingly, the main peak turns out to
be in agreement with what expected from simulations of cluster
dynamical evolution and the ratio between these two radii well correlates
with an empirical dynamical age indicator recently defined from the
observed shape of blue straggler star radial distribution, thus suggesting
that no exotic mechanisms of energy generation are needed in the cores of
the analyzed clusters.
\end{abstract}

\keywords{ catalogs -- galaxies: star clusters: general -- globular clusters: general}

\section{Introduction}
Globular clusters (GCs) represent one of the most intensively
investigated astrophysical systems in the Universe. Indeed, the
comprehension of their origin and nature has implications for
numerous, important fields of Astrophysics and Cosmology, from the
formation of the first self-gravitating objects in the $\Lambda$CDM
cosmological scenario \citep[][see also references
therein]{krav05}, to the theory of stellar evolution and the
formation of stellar exotica (like blue stragglers and millisecond
pulsars), which is made possible by the peculiarly dense and
 dynamically active environmental conditions of these systems
  \citep[e.g.,][]{bailyn93,bellazzini95, ferraro95, rasio07,ferraro09a}.  Their
properties also provide crucial information on the formation and
evolutionary mechanisms of the Galaxy \citep[e.g.,][]{trem75,quinlan,
  ashman98, cdm, ferraro09b, forbes}, as well as on the processes characterizing
the dynamical evolution of collisional systems
\citep[e.g.,][]{mh97,ferraro12}.

Despite the undoubted importance of precisely and accurately determining their
properties, most of the Galactic GC structural and morphological parameters
are still derived from surface brightness (SB) profiles
extracted from mid-80's CCD images and, in a minority of
cases, from star counts on photographic plates mostly dating back to
late 60's--70's \citep{trager95}. Even the most recent parameter
compilations \citep[][hereafter \citetalias{mclaugh}; \citeauthor{wang13} \citeyear{wang13} for M31 clusters]{mclaugh} are based
on SB measurements. 
Indeed, SB profiles are known to suffer  from possible bias due to the presence
of very bright stars \citep[see, e.g.,][for the discussion of methods
  trying to correct for this problem]{ng06}.
Instead, every star has the same ``weight'' in the construction of the
number density profile and no bias is therefore introduced by the
presence of sparse, bright stars.  For this reason, resolved star
counts represent the most robust way for determining the cluster
density profiles and structural parameters \citep[see, e.g.,][]{lug95,
  fe99b, fe03}. In spite of these advantages, however, only a
few studies regarding individual or very small sets of clusters
\citep[e.g.][]{salinas} have been performed to date, while, to
our knowledge, no catalogs of star count profiles sampling the
entire cluster radial extension can be found in the literature.
This is essentially due to the fact that the construction of
complete samples of stars both in the highly crowded central region
and in the outermost part of clusters is not an easy task. 
It requires the proper combination of high-resolution photometry
sampling the cluster centers and high-precision wide-field imaging of
the external parts. In particular, an appropriate coverage of even the
regions beyond the tidal radius is necessary to get a direct estimate
of the level of contamination from background and foreground Galactic
field stars.

It is worth noting that both the inner and the outer portions of the profile provide
crucial information on the structure of the cluster. In fact, the
central part constrains the core radius, the central density, and also
the possible existence of a power-law cusp \citep{ng06,ng07} due to the post-core
collapse state of the system \citep{djorg86,trenti10}, or to the
presence of an Intermediate-Mass Black Hole \citep[IMBH;][but see also
  Vesperini \& Trenti 2010]{BW,baum05, mioc07}.
The external portion provides information on the possible presence of
tidal tails and structures well outside the cluster Roche lobe, that
are indeed observed in a growing number of GCs \citep[see,
  e.g.,][]{leon00,testa,oden03,belo,koch09,jordi10,sollima}.  The
influence of escaped stars (either originated by two-body internal
relaxation or by tidal stripping due to the external field) on the
outer density profile makes more and more questionable the use of the
widely employed \citet{king66} model \citep[see, e.g., the catalogs
  of][ 2010 version, hereafter H10, and \citetalias{mclaugh}]{djorg93,
  pryor93, trager95, harris}.  In this model, the tidal effect is
imposed by construction with a sharp cutoff of the Maxwellian
distribution at the ``limiting radius'', while many clusters seem to
show a radial density that drops towards the background level much
more smoothly than the King model predicts, even following a
scale-free power-law profile \citep[][but see also \citeauthor{williams12}
\citeyear{williams12} for a recently proposed ``collisionless'' model]
{grillmair95,jordi10, kupper10, carballo12, zocchi12}.
For this reason, \citetalias{mclaugh} tested the \citet{wilson75}
model to reproduce the SB profile of Milky Way and Magellanic Clouds
GCs, finding that most of the latter and $\sim 80\%$ of the
Galactic sample are better fitted by this alternative model, which
gives a smoother cutoff at the limiting radius. 
However, this could be due to a not appropriate coverage of the
cluster external region and therefore a not accurate background
decontamination.  Recently, \citet{carballo12} used wide-field star
count data to study the very outer parts of 19 Galactic GCs in the
inner-halo, showing that King and power-law models both provide
reasonable fits to the observations in most of the cases, though the
latter gives a better representation for $\sim 2/3$ of their
sample. Finally, a substantial equivalence of King and Wilson models
in representing the structure of 79 globulars in M31 was found by a very
recent collection of HST SB profiles \citep{wang13}.

In this paper we provide the first homogeneous catalog of star count
density profiles and derived structural parameters, for a
sample of 26 Galactic GCs. Both King and Wilson models are used to fit
the observations.  We specifically focus on apparently ``normal'' GCs,
showing a star count central density with no significant
deviations from a flat behavior (hence no
post core-collapsed systems or clusters with a central density cusp
have been included in the sample).  The paper is organized as follows:
in Sect.~\ref{obs} we give some details about the construction of the
observed density profiles; in Sect.~\ref{models} the adopted
self-consistent models are outlined and defined and a description of
the best-fitting procedure is given; finally, conclusive remarks are
presented in Sect.~\ref{discuss}.

\section{Observed star count profiles}
\label{obs}
In all cases (but the loosest object, NGC 5466), the cluster central
regions have been sampled with high-resolution HST observations, thus
properly resolving stars even in the most crowded environments.  These
data have been combined with complementary sets of wide-field
ground-based observations in order to cover the external parts of
the target clusters, thus sampling the entire radial extension
and, in most of the cases, even beyond \citep[see, e.g.,][and references therein]
{lanzoni07b,lanzoni07c,dalessandro08}.
The projected density profile
of each cluster has been determined from direct star counts in
concentric annuli around the gravity center\footnote{
The measured center is actually not weighted by stellar masses, but
it is simply based on an arithmetic average of star coordinates.
}($C_\mathrm{grav}$).
While the procedure is described in detail in each specific paper (see
references in Table \ref{obstab}), here we quickly summarize the main
steps.

%\clearpage
%\begin{deluxetable*}{lllcl}
\begin{deluxetable}{lllcl}
%\tabletypesize{\scriptsize}
\tablecaption{Centers of gravity
\label{obstab}}
\tablewidth{0pt}
\tablehead{
\colhead{NGC name} & \colhead{$\alpha$} & \colhead{$\delta$} &  \colhead{$\sigma_{\alpha,\delta}$}
& \colhead{Ref.}\\
    & \colhead{(h:m:s)}    & \colhead{($\textrm{deg} : \arcmin:\arcsec$)}     &  \colhead{(\arcsec)}
}
%\hline
\startdata
104 (47Tuc)& $00:24:05.71$ & $-72:04:52.20$& $0.5$  &1  \\ %rifatto 26/04
%NGC\,104 (47Tuc)& $00:24:05.20$ & $-72:04:51$   & $1$  &\citet{ferraro04}  \\
288        & $00:52:45.24$ & $-26:34:57.40$& $1.8$ &1, 2  \\
1851       & $05:14:06.755$& $-40:02:47.47$& $0.1$ &1     \\ %fatto da me
1904 (M79) & $05:24:11.09$ & $-24:31:29.00$& $0.5$  & 3 \\
2419       & $07:38:8.47$  & $+38:52:55.0$ & $0.5$  & 4\\
5024 (M53) & $13:12:55.18$ & $+18:10:06.1$ & $0.5$  &1\\%fatto da me
5272 (M3)  & $13:42:11.38$  & $+28:22:39.1$& $1$    &1\\ % rifatto da Barbara
5466       & $14:05:27.25$  & $+28:32:01.8$& $2$     & 1\\ % rifatto da Barbara
5824       & $15:03:58.637$ & $-33:04:05.90$&$0.2$& 1\\ % fatto da Nicoletta
5904 (M5)  & $15:18:33.214$ & $+02:04:51.80$  & $0.2$ &1\\ % rifatto da Barbara 7/05/13 % \citet{lanzoni07a} \\
%NGC\,??6093 (M80)\\ tolto
6121 (M4)  & $16:23:35.03$ & $-26:31:33.89$ & $1$   &1\\%fatto da B
6205 (M13) & $16:41:41.21$ & $+36:27:35.61$& $0.4$  &1\\%FATTO DA B il 29/04/13
6229       & $16:46:58.74$ & $+47:31:39.53$ & $0.1$ & 5 \\
6254 (M10) & $16:57:8.92$  & $-04:05:58.07$ & $1$   & 6 \\
%NGC\,6266 (M62) &$17:01:12.78$ & $-30:06:46.0$ & $0.5$ &  \citet{beccari06} \\
6266 (M62) & $17:01:12.98$ & $-30:06:49.00$& $0.2$ &  1\\%fatto da B
6341 (M92) & $17:17:07.43$ & $+43:08:09.26$& $0.1$ & 1\\%fatto da B
6626 (M28) & $18:24:32.73$ & $-24:52:13.07$& $0.7$ & 1\\%rifatto da B il 13/05/13
6809 (M55) & $19:39:59.84$  & $-30:57:50.81$   & $1$   & 1\\ %rifatto da B il 6/05/13 
6864 (M75) &  $20:06:4.85$ & $-21:55:17.85$ & $0.5$ & 7\\
7089 (M2)  & $21:33:26.96$ & $-00:49:22.97$ & $1$   &8\\ 
AM 1       & $03:55:02.5$  & $-49:36:53.2$ & $1$ &  9 \\
Eridanus   & $04:24:44.7$  & $-21:11:13.9$ & $1$ &  9 \\
Palomar 3  & $10:05:31.56$ & $+00:04:21.74$& $2$ &  9 \\ %ricontrollato da B il 7/05/13
Palomar 4  & $11:29:16.47$  & $+28:58:22.38$ & $>2$ &  9 \\ %ricontrollato da B il 7/05/13
Palomar 14 & $16:11:00.8$  & $+14:57:27.8$ & $1$ &  10 \\
Terzan 5   & $17:48:04.85$ & $-24:46:44.6$ & $1$ &  11 \\
\enddata
\tablecomments{
Centers of gravity and references for the star count
  surface density profiles of all the GCs in our sample. 
The $\alpha$ and $\delta$ coordinates of $C_\mathrm{grav}$ are
referred to epoch J2000. Their uncertainty (the same in $\alpha$ and
in $\delta$) is given in column 4, in units of arcseconds.
}
\tablerefs{
(1) this work; (2) \citet{gold10}; (3) \citet{lanzoni07c};
(4) \citet{dalessandro08}; (5) \citet{sanna12}: (6)  \citet{ema11};
(7) \citet{contreras12}; (8) \citet{dalessandro09}; (9) \citet{beccari12};
(10) \citet{beccari11}; (11) \citet{lanzoni10}.}
\end{deluxetable}
%\end{deluxetable*}

At odds with many previous studies that adopts as cluster center the
position of the SB peak, for each GC in our sample we
computed $C_\mathrm{grav}$ from star counts,
thus to avoid any possible bias introduced by the presence of a few
bright stars. $C_\mathrm{grav}$ is determined by averaging the
right ascension ($\alpha$) and declination ($\delta$) of all stars lying within a circle
of radius $r$.  Depending on the available datasets and the cluster
characteristics, in every GC we selected the optimal range of stellar
magnitudes, thus to have enough statistics and avoid spurious effects
due to photometric incompleteness (that especially affects the
innermost, crowded regions).  The radius $r$ is chosen as a compromise
between including the largest number of stars and avoiding the gaps of
the instrument CCDs. 
The adopted values of $r$ always exceed the cluster core
radius as quoted by H10, thus to be sensitive to the portion of the
profile where the slope changes and the density is no more
uniform. The search for $C_\mathrm{grav}$ starts from a first-guess
center and stops, within an iterative procedure, when convergence is
reached.  As a further consistency check, the
center was also determined by averaging the stellar coordinates
\emph{weighted} by the local number density, in a way similar to that
outlined by \citet{casertano} in the context of cluster $N$-body
simulations \citep[see][for more details]{lanzoni10}. The two
estimates turn out to be consistent within the errors, as it is indeed
expected in the case of flat-core profiles (like those included in the
present sample). The values of $C_\mathrm{grav}$ adopted in the present paper
are listed in Table \ref{obstab}.

\begin{figure}
%\plotone{kappa.ps}
%\includegraphics[width= 8.3 truecm]{corrNbg.ps}
\includegraphics[width= 8.3 truecm]{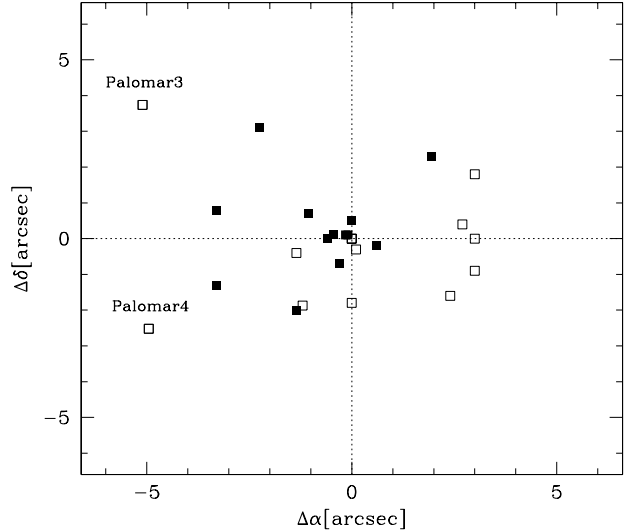}
\caption{Differences between our determination of the coordinates of
each GC center (Table \ref{obstab}) and that quoted in the literature
  \citep[filled squares:][open squares: H10 for GCs not included in the
    \citeauthor{gold10} sample]{gold10}. Reported are the names of
  clusters showing a difference $>4\arcsec$.
%47tuc aggiornato (26/04/13)
}
\label{centers}
\end{figure}

Figure \ref{centers} shows the differences between
the coordinates of the cluster centers of Table \ref{obstab}
and those quoted in \citet{gold10} or in H10 for
those GCs not included in Goldsbury's sample. Differences in both
right ascension and declination are always
smaller than $\sim 4\arcsec$, but for two GCs, namely Palomar 3 and
Palomar 4.  These are two very loose clusters, with extremely low
stellar densities even within the core region. Hence, the
determination of their center is much more difficult, as also
testified by the quite large resulting uncertainties
($\sigma_{\alpha,\delta}\ge 2\arcsec$). In addition, the center of
Palomar 4 quoted in H10 has been determined from scanned plates
\citep{shawl86} and can therefore be inaccurate. In any case, it is
worth noticing that in the case of such loose GCs, with relatively
large core, no significant impact on the density profile is expected
from a few arcsecond erroneous positioning of $C_\mathrm{grav}$.

%\clearpage

The projected number density profile, $\Sigma_*(r)$, is determined by
dividing the entire data-set in $N$ concentric annuli, each one
partitioned in four subsectors (only two or three subsectors are used
if the available data sample just a portion of the annulus).  The
number of stars in each subsector is counted and the density is
obtained by dividing this value by the sector area. The stellar
density in each annulus is then obtained as the average of the
subsector densities, and the uncertainty is estimated from the
variance among the subsectors.  Also in this case, only stars within
a limited range of magnitudes are considered in order to avoid spurious effects
due to photometric incompleteness\footnote{
Note that the considered stars are generally selected over the 
RGB/SGB/TO or the upper MS. Thus, they are fully compatible, in terms
of mass, to the bright RGB that dominates the integrated GC optical emission
from which SB profiles are commonly derived.
}.
As described above, the innermost
portion of the profile is computed by using high-resolution HST data,
while the outer part is obtained from wide-field ground-based
observations. The two portions are normalized by using the annular regions
not affected by incompletness that are in common between the two data-sets.

The observed stellar density profiles are shown in Figure \ref{profs}
(open symbols) for the 26 GCs in the sample\footnote{All the observed profiles
are publicly available at the web site\\ \url{http://www.cosmic-lab.eu/Cosmic-Lab/Products.html}}.
In most of the cases
the collected dataset covers the entire cluster extension, reaching
the outermost region where the Galactic field stars represent the dominant
contribution with respect to the cluster. The spatial distribution of 
field stars is approximately uniform on the considered radial bin scales, and
this produces a sort of ``background plateau'' in the outermost region of the
star count profile. Hence, by averaging the values of the $N_\mathrm{BG}$
points in this plateau, we estimate the
Galaxy background contamination to the cluster density (short-dashed lines
in Figure \ref{profs}). The decontaminated cluster profile, obtained
after subtracting the Galaxy background level, is finally shown
as black symbols in the figure.  As apparent, after the field
subtraction, the profile remains unchanged in the inner and most
populous regions, while the cluster data points can be significantly
below the background level in the most external parts. As a
consequence, the accurate measure of the background level is crucial
for the reliable determination of the outermost portion of the
profile.

%\clearpage
\setcounter{figure}{1}
\begin{figure*}
\center
%\plotone{kappa.ps}
\includegraphics[width= 16.5 truecm]{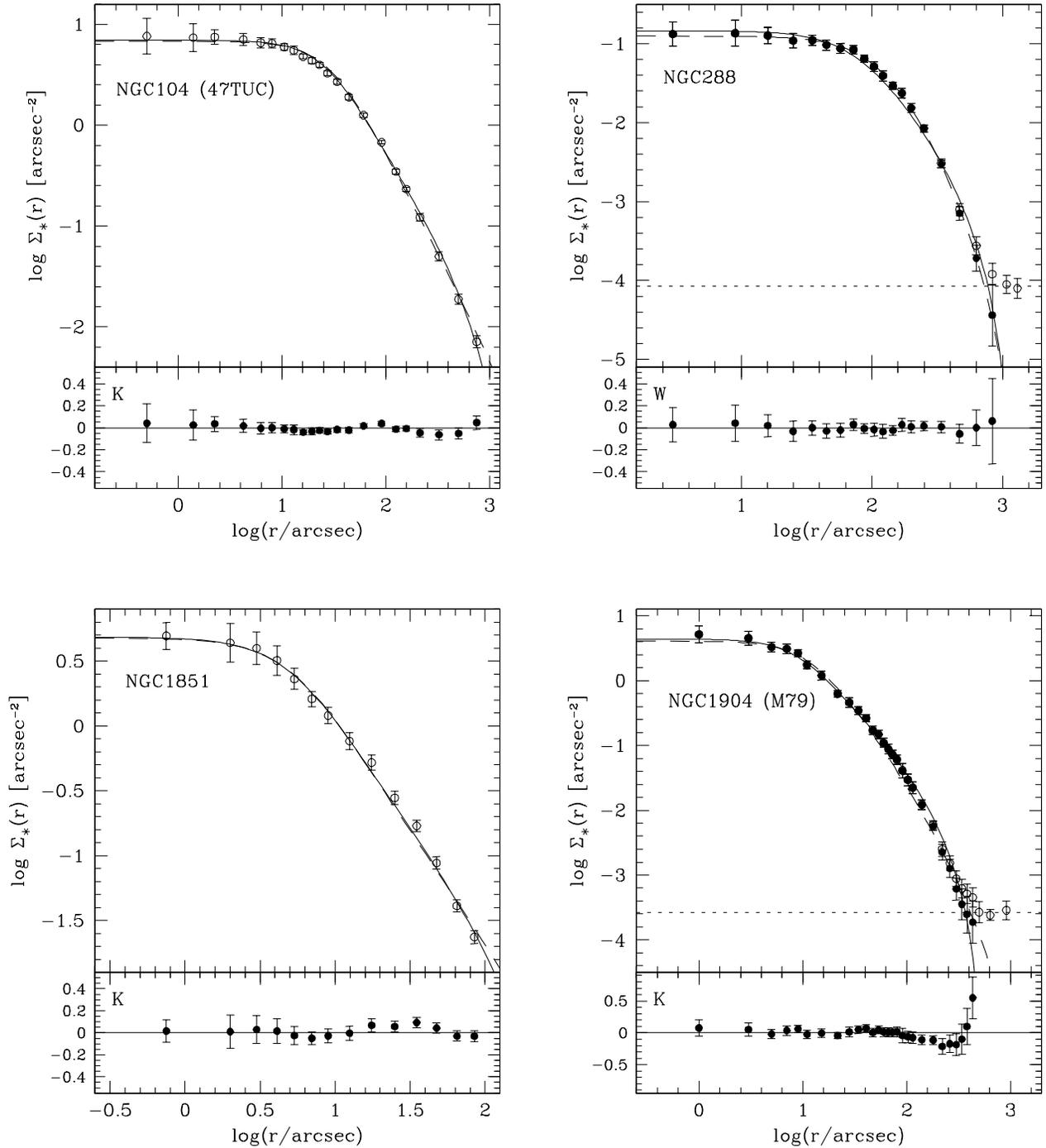}
\caption{Observed star count profiles and corresponding best-fit King
(solid curve) and Wilson (long-dashed curve) models. 
For each cluster, in the upper panel the open circles mark the
observed star count surface density profile, while solid circles
correspond to the profile after the subtraction of the the Galactic field
background density estimate (short-dashed line, if available).
The lower section of each panel shows the residuals between the
(decontaminated)
observed profile and the model (K=King, W=Wilson)
with the lowest value of $\chi^2_{\nu}$.
The error bars of the decontaminated points include the uncertainty in
the background determination.
}
\label{profs}
\end{figure*}

%\clearpage
\setcounter{figure}{1}
\begin{figure*}
\center
%\plotone{kappa.ps}
\includegraphics[width= 16.5 truecm]{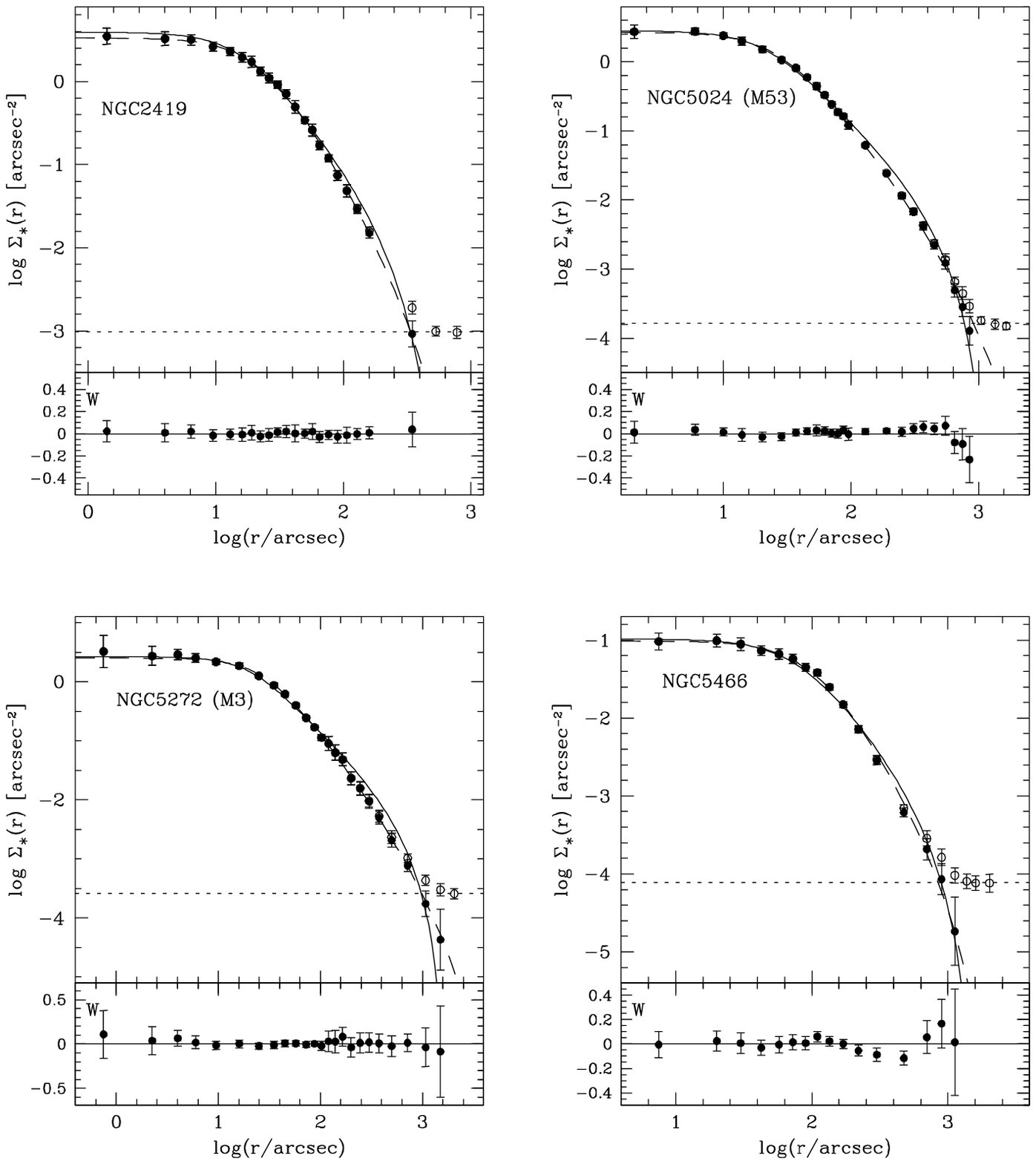}
\caption{(Continued)}
\end{figure*}

%\clearpage
\setcounter{figure}{1}
\begin{figure*}
\center
%\plotone{kappa.ps}
\includegraphics[width= 16.5 truecm]{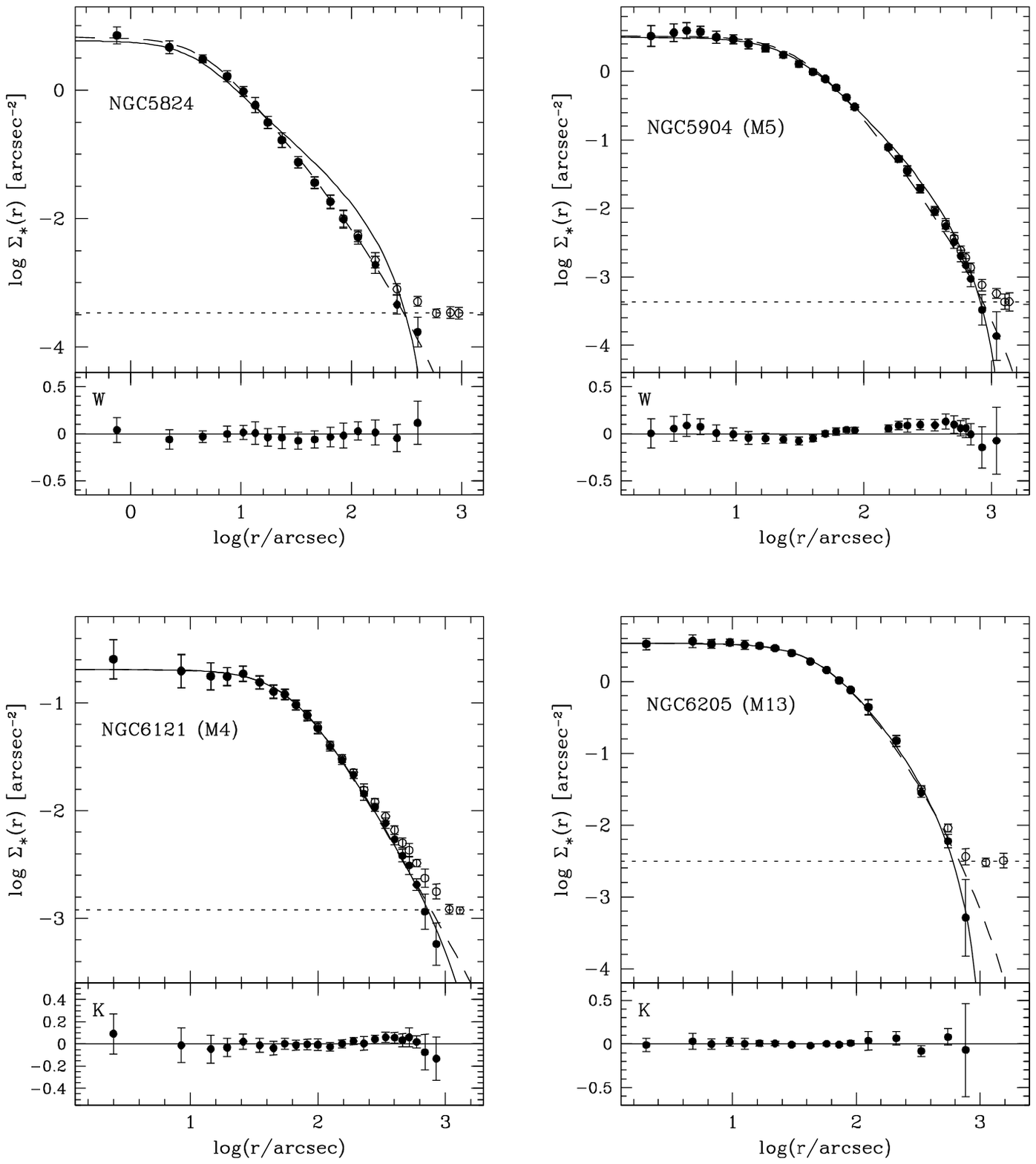}
\caption{(Continued)}
\end{figure*}

%\clearpage
\setcounter{figure}{1}
\begin{figure*}
\center
%\plotone{kappa.ps}
\includegraphics[width= 16.5 truecm]{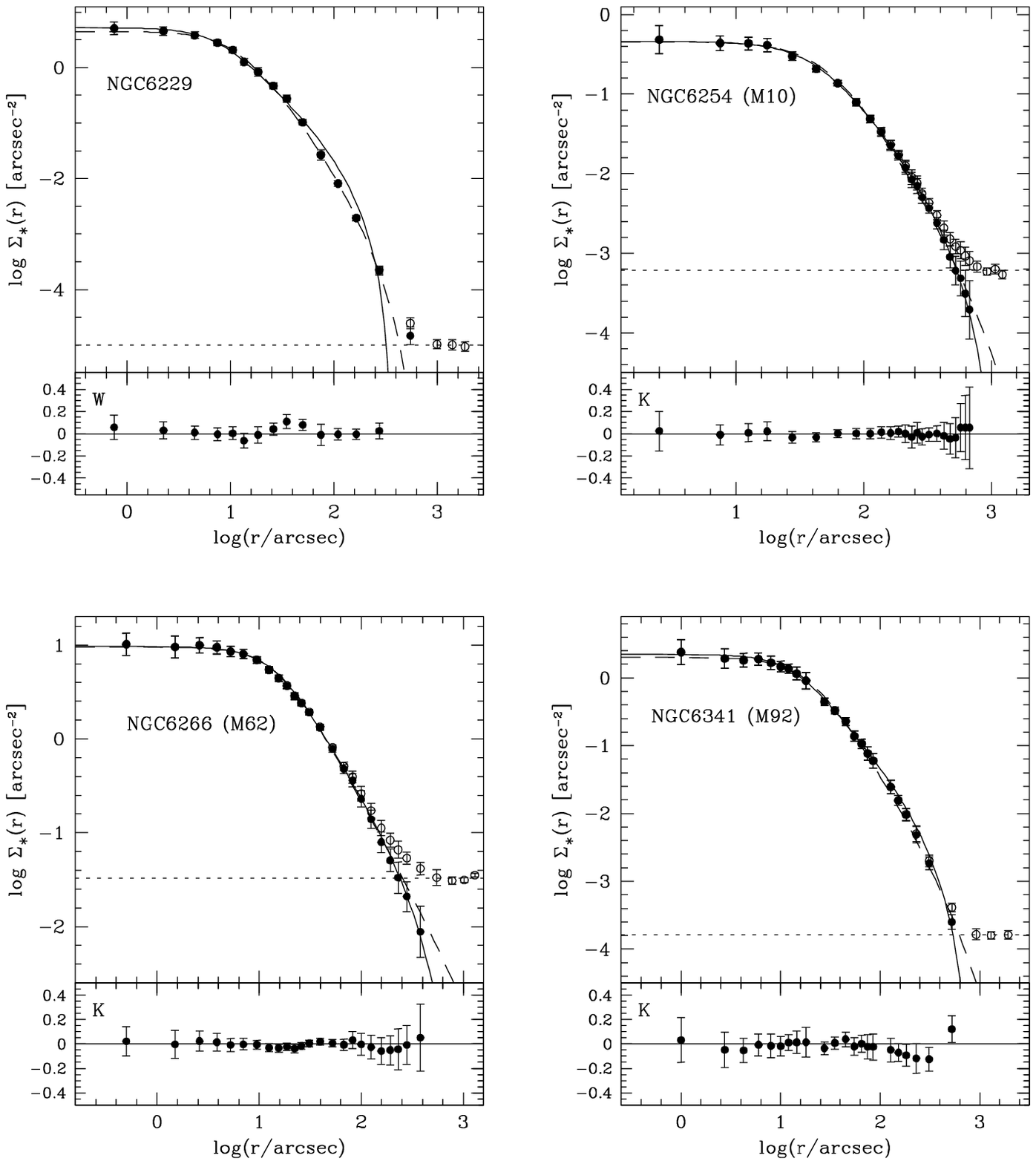}
\caption{(Continued)}
\end{figure*}

%\clearpage
\setcounter{figure}{1}
\begin{figure*}
\center
%\plotone{kappa.ps}
\includegraphics[width= 16.5 truecm]{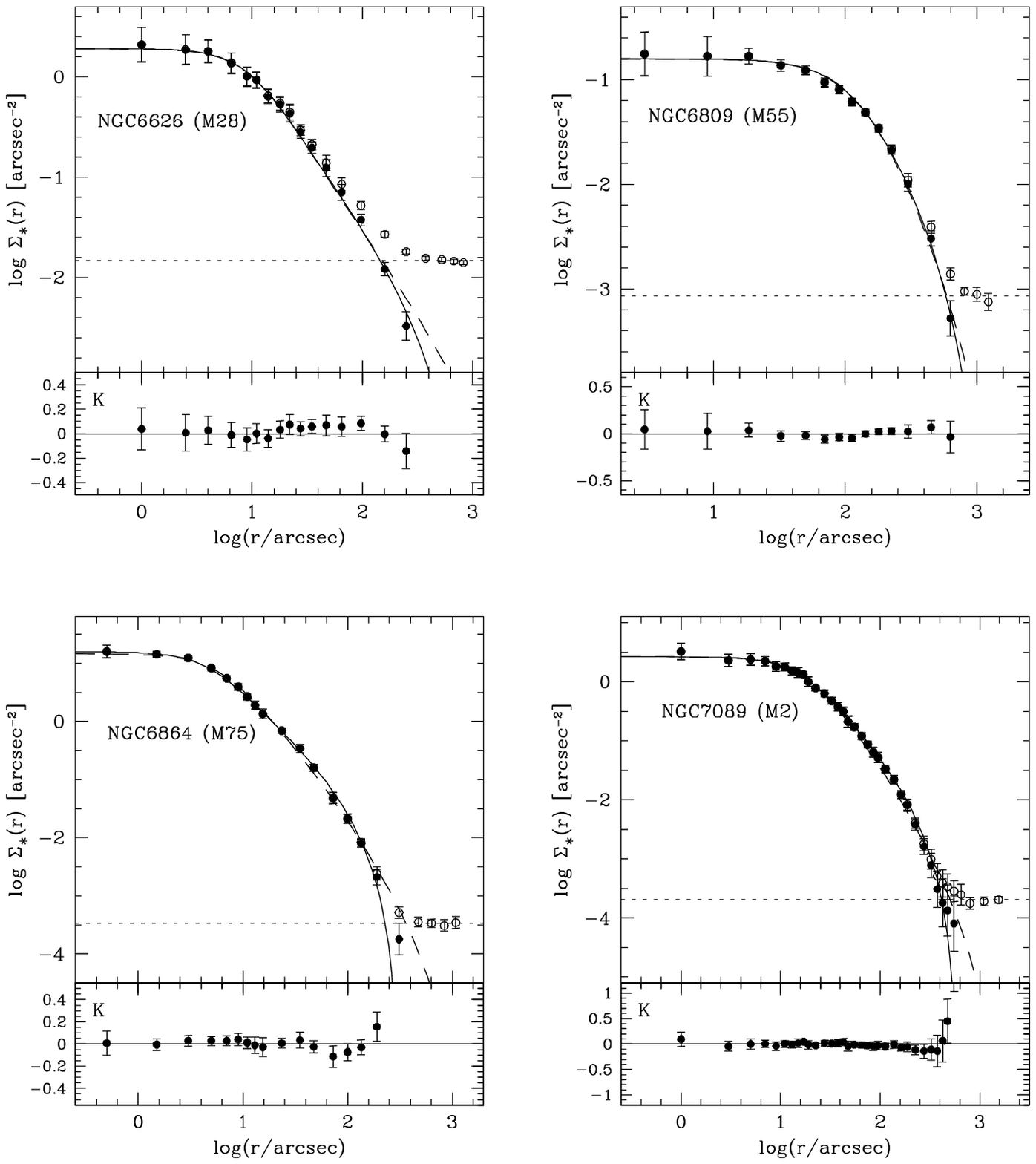}
\caption{(Continued)}
\end{figure*}

%\clearpage
\setcounter{figure}{1}
\begin{figure*}
\center
%\plotone{kappa.ps}
\includegraphics[width= 16.5 truecm]{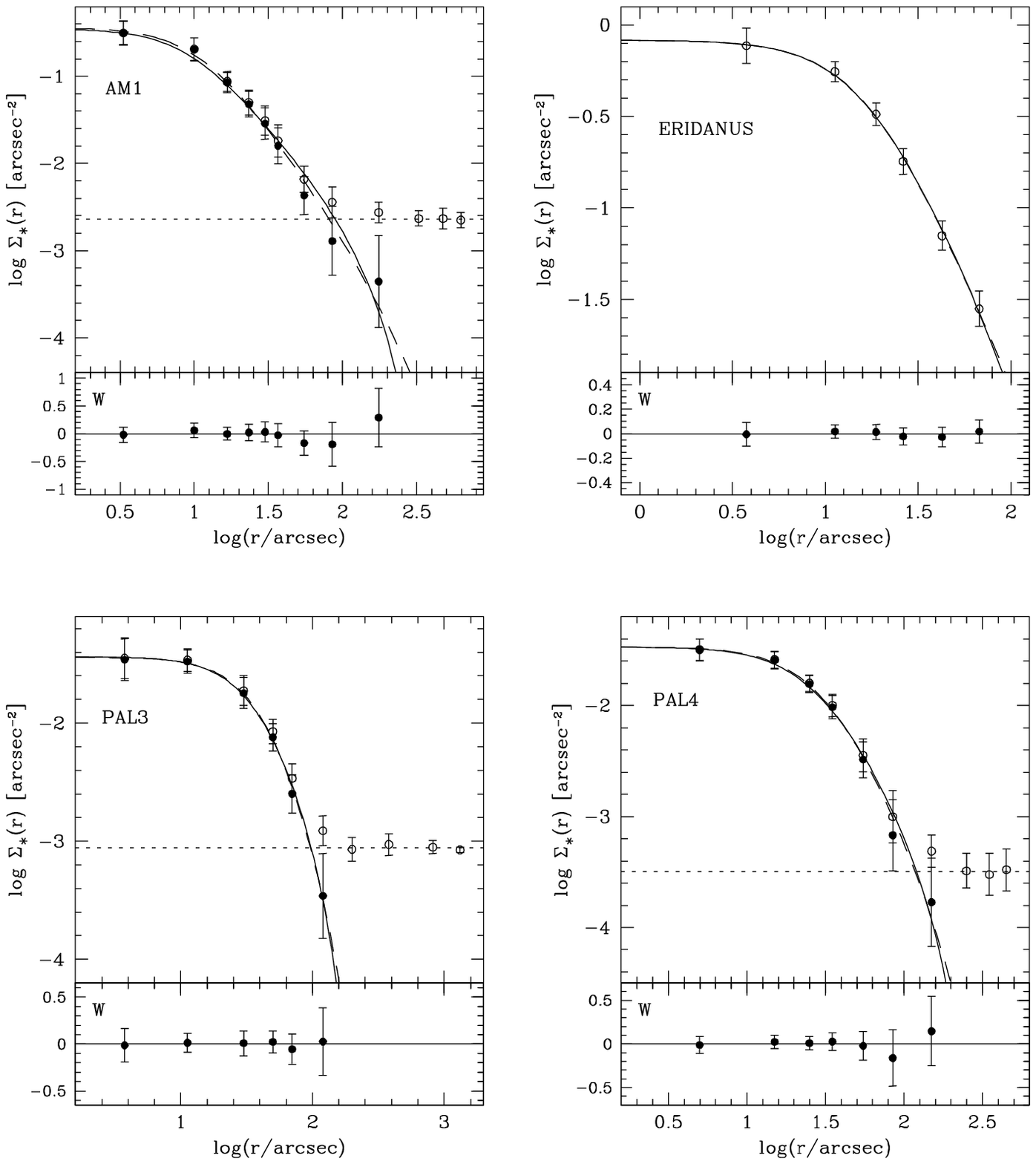}
\caption{(Continued)}
\end{figure*}

%\clearpage
\setcounter{figure}{1}
\begin{figure*}
\center
%\plotone{kappa.ps}
\includegraphics[width= 16.5 truecm]{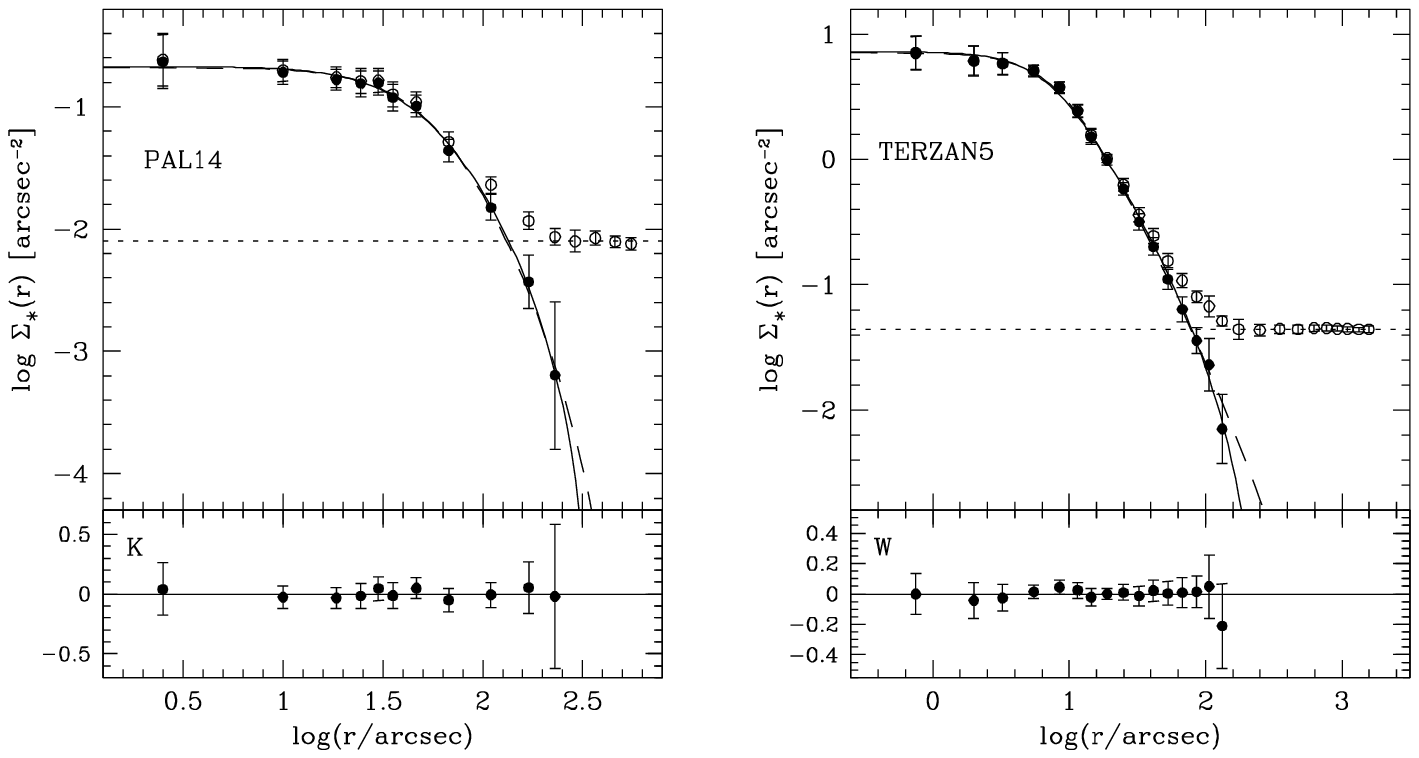}
\caption{(Continued)}
\end{figure*}

\section{Models}
\label{models}
To reproduce the observed star density profiles and thus to derive the
cluster structural parameters, we considered both the
\cite{king66} and the \cite{wilson75} models \citep[see
  also][]{hunter77}, in the isotropic, spherical and single-mass
approximation.  These models (the former, in particular) have been
widely used to represent stellar systems like GCs, that are thought to
have reached a state of (quasi-)equilibrium similar to the one attained by
gases following the Maxwellian distribution function.
Besides the generally good agreement with observations 
\citep[but see the discussion
in][]{williams12} and its valid physical motivations,
the King model has been also derived from a rigorous statistical mechanics
treatment \citep{madsen96}.

Qualitatively, the projected density profiles of the King and Wilson
models are characterized by a constant value in the innermost part
(the ``core''), and a decreasing behavior outwards, with the Wilson
model showing a more extended outer region \citepalias[see Appendix \ref{app1}
and, e.g., Figures 9 and 10 in][]{mclaugh}.
In both cases the density profiles constitute a \emph{one}-parameter
family. This means that the profile \emph{shape} is uniquely
determined by the dimensionless parameter $W_0$, which is proportional
to the gravitational potential at the center of the system. In
practice, the higher $W_0$ the smaller is the cluster core with
respect to the overall size of the system. More details about these
models are presented in Appendix \ref{app1}.

Several characteristic scale-lengths can be defined in both model
families. Some of them have a precise theoretical definition, but no
observational correspondence; some other are commonly adopted in
observational studies, but suffer from some degree of arbitrariness
when measured from the available data.
However, since numerical models have become increasingly more realistic,
much attention has to be paid to give clear and unambiguous definitions
of these parameters so as to allow a closer and
meaningful comparison between theoretical and observational results
\citep[see e.g.][]{hurley07,trenti10}.

Here we consider a number of
different scale radii, thus to allow the widest possible use and the
connection between theory and observations.  We call ``scale radius''
($r_0$) the characteristic length parameter of the model, which most
authors refer to as ``King radius'' in the case of the King
family. This must not be confused with the ``core radius''
($r_\mathrm{c}$), which is operatively defined as the radius at which
the \emph{projected} stellar density $\Sigma_*(r)$ drops to half its
central value (in other studies the SB is considered
instead of $\Sigma_*$).  The values of the scale and core radii are
similar, their difference tending to zero for $W_0\to \infty$.

We define the ``half-mass radius'' ($r_\mathrm{hm}$), as the radius of
the sphere containing half of the total cluster mass.  Of course,
$r_\mathrm{hm}$ cannot be directly observed and we therefore consider
also the ``effective radius'' ($r_\mathrm{e}$), commonly defined as
the radius of the circle that \emph{in projection} includes half the
total integrated light.  In the case of star counts (instead of
SB) profiles, the total integrated light corresponds
to the integral of the number density profile over all radii (i.e. the
total number of observed stars).
For various reasons (including that, with respect to other characteristic
scale-lengths like $r_\mathrm{c}$, it weakly varies during the cluster evolution)
this radius is commonly adopted to measure the GC size \citep[see,
 e.g.,][]{spitz72,lightman78,murphy90}.

Finally, the ``limiting radius'' ($r_\l$) is the model cutoff
radius, at which the density $\Sigma_*$ goes to zero.  This is often
and rather improperly called the ``tidal radius'', even if it is not
directly and trivially related to the tidal effect of the Galactic
field \citep[see also][]{BT, kupper10}.  The logarithm of the ratio between the
limiting and the scale radii is called the ``concentration
parameter'', $c\equiv\log(r_\l/r_0)$. In the considered models
there is a one-to-one relation between the value of $W_0$ and that of
$c$, with the cluster concentration increasing as $W_0$
increases (see Figure \ref{apprel}). Since the Wilson model shows a more extended outer
region, the half-mass, effective and limiting radii, and, as a
consequence, also the concentration parameter, are appreciably larger
than in the King model for any fixed scale radius $r_0$
\citepalias[see Appendix \ref{app1}, and, e.g.,][]{mclaugh}.

\subsection{Best-fitting procedure}
\label{bestfit}
The search for the best-fit to the observed surface density profiles
is performed by exploring a pre-generated grid of $n$ models with the
shape parameter $W_0$ ranging from 1 to 12 and stepped by $0.05$, both
in the King and in the Wilson cases.\footnote{These models can be
generated and freely downloaded from the Cosmic-Lab web site at the
address:\\ \url{http://www.cosmic-lab.eu/Cosmic-Lab/Products.html}.
For each model, the user can also retrieve the line of sight velocity
dispersion profile. In addition, models including a central
intermediate-mass black hole \citep[built by following][]{mioc07} are
also available.}
The corresponding concentration parameters vary between 0.5 and 2.74
in the King case, and between 0.78 and 3.52 for the
Wilson model. The model density profiles are finely sampled in
radius: about 100 logarithmically spaced bins are used, so that linear
interpolation yields accurate estimates at any radius. In order to fit
a given observed (and background decontaminated) profile, the entire grid of models
is scanned and for each value of $W_{0,i}$ (with $i=1,n$) a direct searching algorithm
finds the two scaling
parameters $r_{0,i}$ and $\Sigma_{*,i}(0)$ that minimize the
sum of the unweighted squares of the residuals
and evaluates the corresponding
$\chi^2$ value ($\chi^2_{\mathrm{min},i}$).
At the end of the procedure, the
best-fit model is defined as the one corresponding to the lowest value
among all the obtained $\chi^2_{\mathrm{min},i}$ (let us indicate it
as $\chi^2_\mathrm{best}$).

Results are listed in Table \ref{bigtab}. For each cluster, both the
King and the Wilson best-fit models are given.  The quality of the fit
is reported in the third column of the table in terms of the reduced
$\chi^2$ ($\chi^2_\nu$), i.e. the value of $\chi^2_\mathrm{best}$
divided by the number of the fit degrees of freedom. This is equal to
$N-N_\mathrm{BG}-3$, where $N$ is the total number of observed
points, $N_\mathrm{BG}$ is the number of points used to evaluate the
Galaxy background contamination (see Sect. \ref{obs}), and 3 quantities (two
scale parameters and $W_0$) are evaluated by the best-fit. As apparent,
for many GCs the two models give an approximately equivalent fit to
the data, while in a few cases the observations are significantly
better reproduced by one of the two (see Sect.\ref{discuss}).

The 1-$\sigma$ confidence intervals of the best-fitting parameters
(see Table \ref{bigtab}) are estimated from the distribution of the
$\chi^2_{\mathrm{min},i}$ values, in line with the method of the
$\Delta \chi^2$ described, e.g., in \citet{recipes}. From this
distribution we select the sub-set of models with
$\chi^2_{\mathrm{min},i} \le \chi^2_\mathrm{best} + 1$. Then, the
1-$\sigma$ uncertainty range of a parameter is assumed to be equal to
the maximum variation of that parameter within this sub-set of models
\citepalias[as it is done in][]{mclaugh}.  In some cases, this
procedure yields large uncertainty ranges either because the fit is
not very good (for example, in the case of the King model fit of NGC
2419), or because there is a relatively small amount of data points
(as in the cases of Palomar 3 and Palomar 4).  Moreover, as can be
appreciated in Table \ref{bigtab}, the uncertainty limits are often
asymmetric with respect to the best-fit value.

Given the importance of a correct evaluation of the
Galactic background for a proper definition of the cluster density
profile (Sect.~\ref{obs}), we tried to estimate the sensitivity of
the fitting procedure to this quantity. In general, we found that a
change in the background level can significantly affect only the
one/two most external points considered in the fit procedure. This can
possibly change the best-fit value of $W_0$ (and hence of $c$), while
the scale radius is essentially unaffected.  However in most of the
cases the large radial coverage of our data-sets guarantees a solid
evaluation of the background level, and only in a few clusters (see
footnotes in Table~\ref{bigtab}) the exclusion of the last data point allows a
considerable improvement of the fit, possibly suggesting that the
Galactic background could be underestimated for these systems.

\section{Discussion}
\label{discuss}
Figure \ref{profs} shows the observed density profiles and the
results of the fitting procedure for all the program clusters. 
The best-fit King and Wilson profiles are plotted as
solid and long-dashed lines, respectively. The lower panel shows the
residuals with respect to the model that provides the lowest value of
$\chi^2_\nu$, and,
in the following, we call K-type (or W-type) clusters those for which
this model is the King (or Wilson) one.
Our analysis classifies $50\%$ of GCs in our sample as W-type. %13=K, 13=W
This percentage is smaller than that found by \citetalias{mclaugh}
for their Galactic sample, but the different size of the two samples should be
taken into account. In fact, a change of classification for just a few
clusters in our case would suffice to significantly alter the overall
percentage.

The collected catalog offers the possibility of a
meaningful comparison with results obtained from SB profiles. We
noted, for instance, that three clusters in common with our sample
(namely NGC 1851, M13 and M62) show hints of a SB central cusp in the
work of \citet{ng06}. No evidence of such a feature is instead found
in the star density profile shown in Fig.~\ref{profs}. A close inspection of the
SB profiles published in \citet{ng06} reveals that, although their
data sample a region more internal with respect to ours, a deviation
from a flat core behavior should be already appreciable in the region
sampled by our observations, at least in correspondence to our
innermost data point. Indeed this disagreement could be the
manifestation of the typical bias affecting the SB profiles, where a
group of a few bright giants can produce a spurious enhancement of the
SB, not corresponding to a real overdensity of stars.

Comparing our results with those presented by \citetalias{mclaugh} for
the 23 clusters in common (i.e., all clusters in our sample, but
 NGC 6626, Eridanus and Terzan 5), we find that the same type of
best-fit model is obtained for 15 GCs, five (ten) of which are
best fitted by a King (Wilson) model in both studies.  For the
remaining eight GCs the two works provide different best-fit
model (seven are K-type in our study and W-type in \citetalias{mclaugh},
and vice-versa for the remaining one), at least formally.  Indeed,
significant differences ($\gtrsim 30\%$) between the quoted $\chi^2_\nu$
values are found only for two clusters, namely M2 and M10, which are of
K-type in our study and of W-type in \citetalias{mclaugh}.  A detailed
inspection of Figure \ref{profs} shows that the discrimination between the
two types of model adopted here is often driven by the last,
background-subtracted, points. On the other hand, the datasets used
by \citetalias{mclaugh} are less radially extended (see their Figure
12) and the last points of their SB profiles are not corrected for the
Galactic background level.  Therefore, we can reasonably state that the
aforementioned differences in the best-fit model classification can be
ascribed to a residual background contamination of the
\citetalias{mclaugh} profiles.

\begin{figure*}
\center
\includegraphics[width= 16.5 truecm]{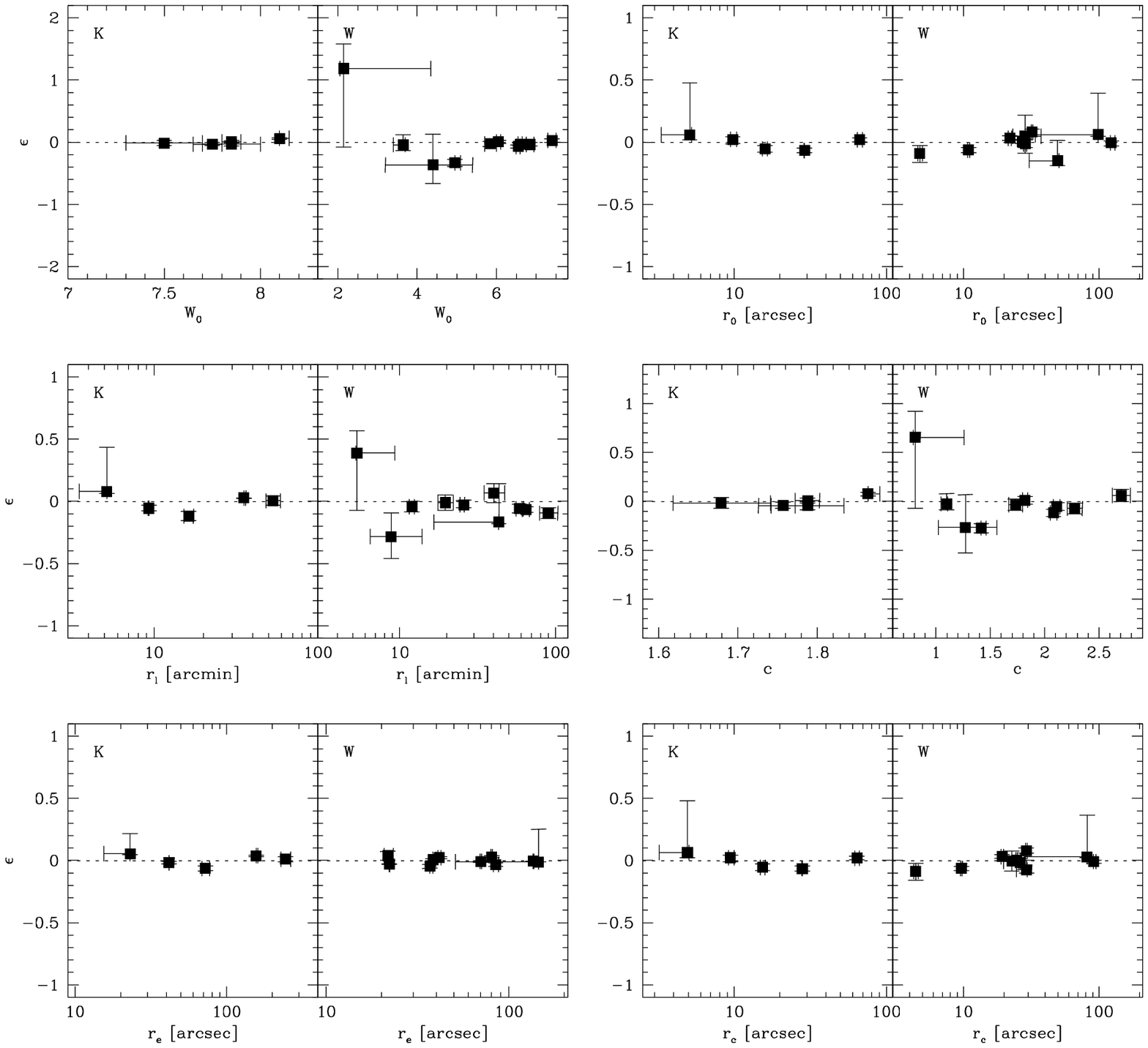}
\caption{Comparison between various best-fit structural parameters as
  obtained in our study and in \citetalias{mclaugh}, for the 15
  clusters that are best-fitted by the same type of model in both
  works. Results for the five K-type GCs and the ten W-type GCs are
  shown, respectively, in the left-hand and in the right-hand sides of
  each panel (see labels). From top-left to bottom-right the
  considered structural parameters are: $W_0$, $r_0$ $r_\l$,
  $c$, $r_\mathrm{e}$, and $r_\mathrm{c}$ (see x-axis labels).  The
  relative difference $\epsilon \equiv (\hat p- p)/p$ between the value
  of the generic parameter quoted by \citetalias{mclaugh} ($\hat p$) and the corresponding
  value obtained in our analysis ($p$) is plotted as a function of $p$.
  %Error bars on $\epsilon$ reflect the given uncertainties on both $p$ and $\hat p$.
  }
\label{confr}
\end{figure*}

\begin{figure*}
\center
\includegraphics[width= 16.5 truecm]{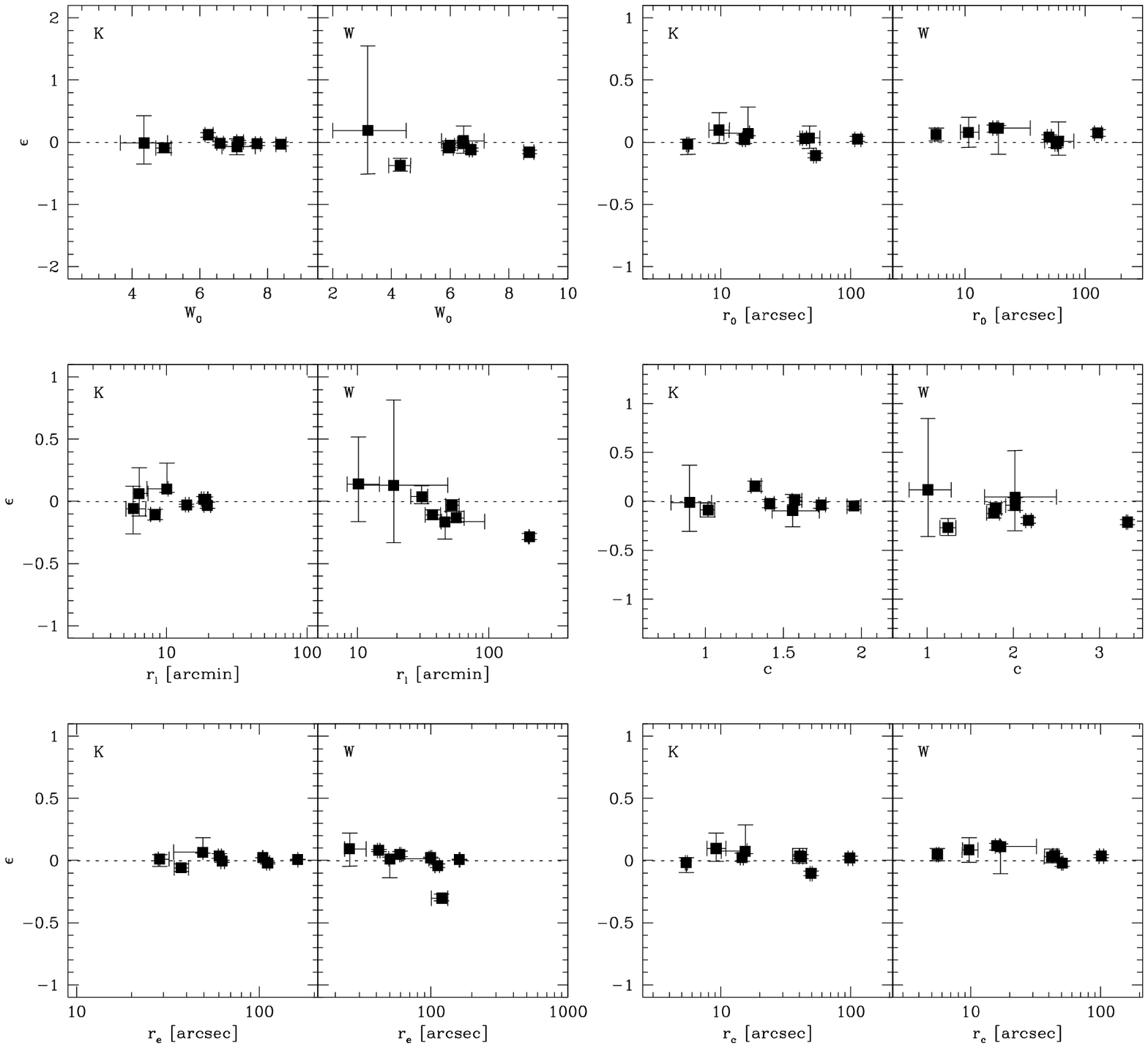}
\caption{The same as in Fig. \ref{confr}, but for the eight clusters
  which are have different type of best-fit model in the two studies:
  the structural parameters corresponding to the King model fitting
  are shown in the left-hand side of each panel, those from the Wilson
  fitting are shown in the right-hand sides.}
\label{confr3}
\end{figure*}

In Figures \ref{confr} and \ref{confr3} we compare some relevant
structural parameters derived in our study and in
\citetalias{mclaugh}.  Fig.~\ref{confr} refers to the 15 clusters for
which the best-fit model is of the same type in both studies, the
left- and right-hand panels concerning, respectively, the five K-type and
the ten W-type GCs. Fig.~\ref{confr3}, instead, refers to the eight
clusters for which the best-fit model is of different type: in the
left-hand panels we compare the structural parameters obtained from King models,
while the right-hand panels refer to Wilson models.  In general, a
good agreement is found between the parameters derived in
our work and in \citetalias{mclaugh}. The largest differences are
found mostly for the the Wilson best-fit limiting radius, which, in turn,
also affects the values of $c$ and $W_0$.  These differences are reasonably
expected, since the Galactic background seems not well sampled in \citetalias{mclaugh}
for most of the clusters in common.  This could explain why the majority of our
estimates of $r_\l$ (and, in turn, of $W_0$ and $c$) are larger
than those quoted in \citetalias{mclaugh}, especially for Wilson best-fit
profiles (because of the more extended envelope).

%\clearpage

%\clearpage

In order to provide a quantitative estimate of the ability of the fitting
procedure to clearly discriminate between the two King and Wilson models, we used the
relative difference between the reduced
$\chi^2$, i.e. the quantity $\Delta \equiv(\chi^2_\mathrm{W}-\chi^2_\mathrm{K})/(\chi^2_\mathrm{W}+\chi^2_\mathrm{K})$
 \citepalias[as in][]{mclaugh}.
Thus, $\Delta= 0$ indicates that the two models provide fits of the
same quality, while $\Delta\simeq 1$ means that the King fit is
substantially better than the Wilson one, and vice-versa for
$\Delta\simeq -1$.

\begin{figure}
%\plotone{kappa.ps}
%\includegraphics[width= 8.3 truecm]{corrNbg.ps}
\includegraphics[width= 8.6 truecm]{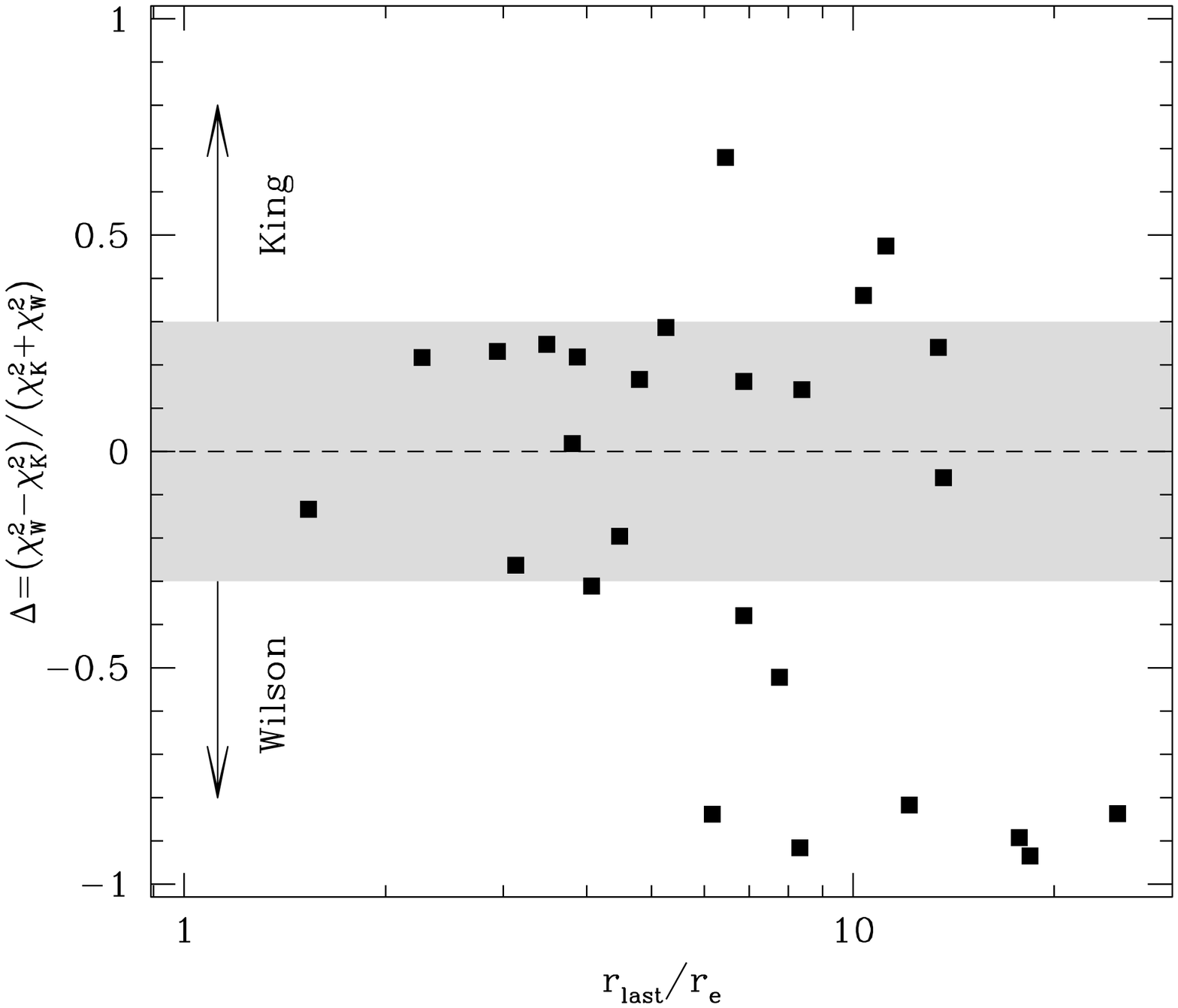}
\caption{
Goodness of the Wilson fit with respect to the King fit (expressed as the relative
difference between the respective $\chi^2_\nu$) for our GC sample
as a function of the observed radial extent of the cluster (see text).
Points with $\Delta>0$ correspond to GCs better fitted by King models, those with $\Delta<0$ are
of W-type. The shaded region (where $|\Delta| \leq 0.3$) includes clusters
whose W- and K-type best-fit profiles turn out to be practically equivalent
at a ``visual inspection''. 
\label{corrBL}}
\end{figure}

In Fig.~\ref{corrBL} we plot $\Delta$ as a function of
$r_\mathrm{last}/r_\mathrm{e}$, where $r_\mathrm{last}$ is the radius
of the outermost point of the decontaminated density profile.  The
quantity along the abscissa is a measure of how many effective radii
are sampled by the observations. Note that in our catalog all clusters
(but 47Tuc, NGC 1851 and Eridanus) are sampled out to where the
Galactic field becomes dominant.  Hence $r_\mathrm{last}/r_\mathrm{e}$
is a measure of the actual extension of each cluster (apart from the
three aforementioned clusters for which it represents only the radial
extension sampled by the observations).  As in \citetalias{mclaugh}
(their Fig. 14, bottom-right panel), we find that the discrimination
becomes more solid in more extended clusters (for
$r_\mathrm{last}/r_\mathrm{e}\gtrsim 6$).  This is indeed expected,
since the King and Wilson model profiles differ only in the external
regions. Interestingly, however, K-type clusters are found also for
large values of $r_\mathrm{last}/r_\mathrm{e}$, thus indicating that
the classification in King or Wilson type is linked to intrinsic
properties of the systems and not due to observational biases (like an
insufficient radial sampling of the profile).  Finally, a hint of a
trend toward best-fit Wilson models ($\Delta\sim -1$) for increasing
radial extension of the cluster seems to be present, but the number of
clusters in our sample is too small to draw a firmer conclusion
concerning this trend. As to this point, it is also important to
notice that King models have an intrinsic upper limit of $\sim 13$ for
$r_\mathrm{last}/r_\mathrm{e}$ (see Fig.\ref{apprel}, second panel
from the bottom, and note that $r_\mathrm{last}\leq r_\l$ by
definition), while Wilson models allow to fit clusters characterized
by larger values of this ratio ($\lesssim 145$).  

%\clearpage

However, it is interesting to note that the large majority of the
cluster lies around $\Delta=0$, thus indicating a not significant
difference in the quality of the fit between the two kinds of models.
To be conservative, we have highlighted as a
gray strip a region\footnote{
This range has been chosen somewhat empirically and it is meant
to be a general guide rather than a rigorous statistical measure.}
($-0.3<\Delta<0.3$) where the $\Delta$ parameter
does not allow a clear-cut preference in the fitting procedure for
either King or Wilson models, in the sense that, by a visual inspection,
they fit equally well the profile, especially in the inner part.

\begin{figure}
%\plotone{kappa.ps}
\includegraphics[width= 8.6 truecm]{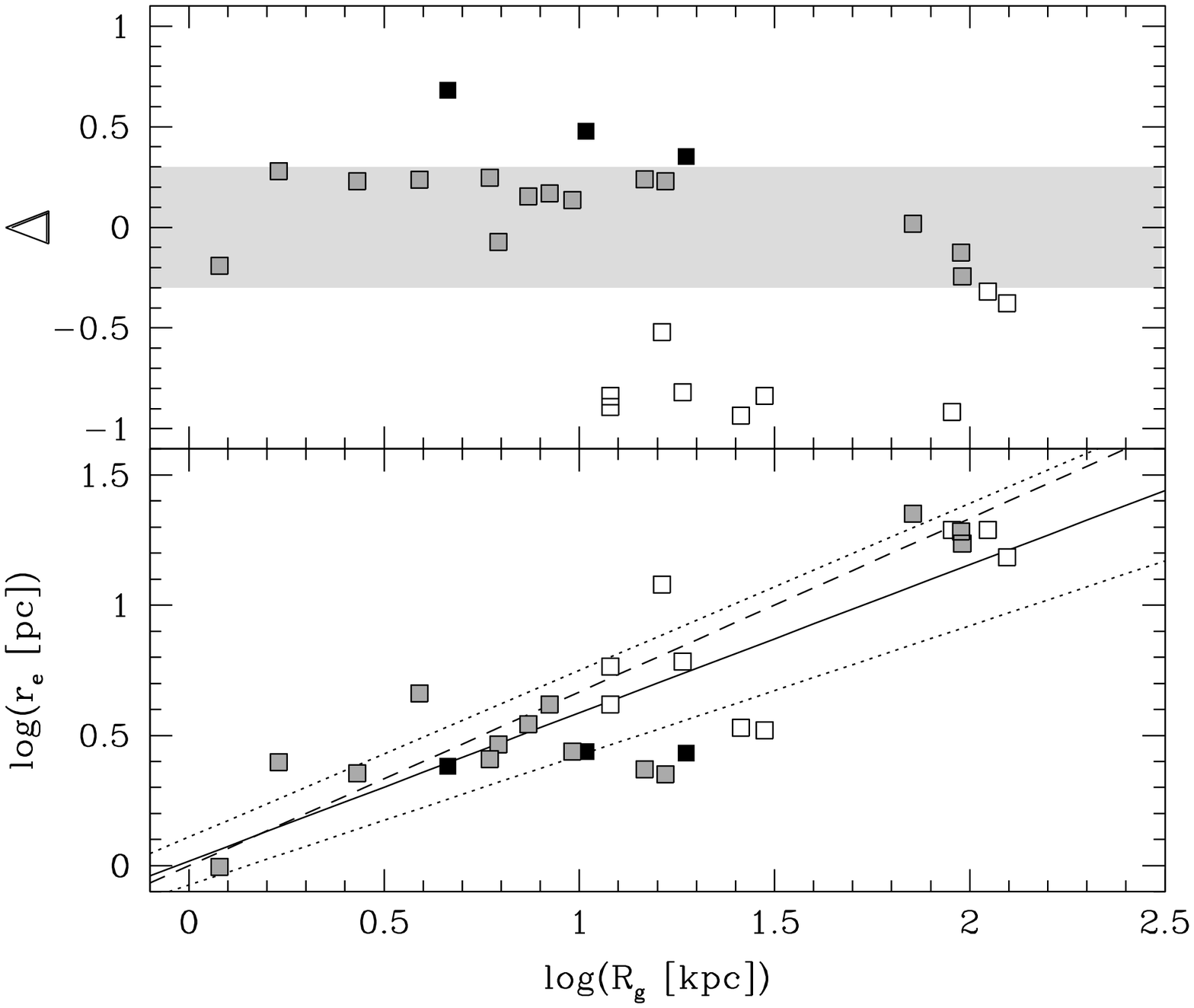}
\caption{Upper panel: the $\Delta$ parameter of Fig.~\ref{corrBL} plotted as a function of the galactocentric
distance of each cluster.
Lower panel: clusters size (effective radius) vs. galactocentric distance; the solid line
corresponds to the linear least
squares fit $\log (r_\mathrm{e}/\mathrm{pc}) = 0.57\log
(R_\mathrm{g}/\mathrm{kpc})+0.018$, with the uncertainty region enclosed by the two
dotted lines  \citep[the dashed line is the $R_\mathrm{g}^{2/3}$ relation; see][]{vdb11}.
Solid and open dots are clusters clearly best-fitted
by King and Wilson model, respectively, while gray squares mark
clusters of ``equivalent'' type classification ($|\Delta| \leq 0.3$, the gray strip in the upper
panel).
\label{corr}}
\end{figure}

The top panel of Fig.~\ref{corr} shows the $\Delta$ parameter as a
function of the galactocentric distance ($R_\mathrm{g}$).
The values of $R_\mathrm{g}$ have been taken from H10, while the
distance moduli are from \citet{fe99a}, with the exception of Terzan 5
\citep[for which we adopted the distance quoted by][]{valenti07} and
all GCs not included in these works (for which the distances quoted in
H10 have been assumed).
According to the discussion above, we have highlighted the clusters with an ``equivalent''
classification as gray squares. Even with these caveats, we notice that
there is a group of clusters between $10$ and $30$ kpc for which Wilson models can fit the
data definitely better than King models.  In the same range of
galactocentric distances there are also clusters best fitted by King
models and clusters for which the two models provide fits of similar
quality. Different orbital properties (and the ensuing differences in
the cluster dynamical evolution) might be responsible for the
existence of these different groups of clusters.

%\clearpage

The analyzed sample has been used also to test the existence of
a relation between the cluster size (usually measured with
$r_\mathrm{e}$) and the galactocentric distance
\citep[see, e.g.,][and references therein]{vdb11,madrid}.  
The result is shown in the bottom panel of Fig. \ref{corr},
where the $r_\mathrm{e}$ values are taken from the type of model giving
the lowest $\chi^2_\nu$. A
correlation is visible (the Pearson's coefficient is
$\simeq 0.84$), with a slope slightly smaller than $2/3$ as derived by
\citet[][using data from the H10 catalog]{vdb11}, but still compatible
within the uncertainties.  The scaling relation we obtain is
\begin{equation}
\label{relation}
\frac{r_\mathrm{e}}{\textrm{pc}}\sim (1.0\pm
0.2)\left(\frac{R_\mathrm{g}}{\textrm{kpc}}\right)^{0.57\pm 0.07}.
\end{equation}
A qualitatively similar trend has been recently observed also for a sample of GCs in
M31 \citep{wang13}.
The importance of the role played by the external Galactic field in determining this
relation \citep[e.g.][]{vdb94}
has been recently suggested on more rigorous theoretical grounds \citep{ernst13}.
 
Following \citet{vdb11,vdb12}, we also tested the existence of a few relations between
cluster structural properties that might also help explaining the observed
scatter around the relation expressed by Eq.~(\ref{relation}). 
In particular, we studied the behavior of
$r_\mathrm{e}$, $c$ and the metallicity [Fe/H] as a function of both
the total absolute magnitude $M_V$ and the parameter $0.57
\log R_\mathrm{g} -\log r_\mathrm{e}$ quantifying the deviations from Eq.~(\ref{relation}). 
We adopted the integrated magnitudes quoted by H10, while 
%distance moduli,
reddening parameters and metallicities have been taken from \citet{fe99a},
or from H10 for the GCs not included in that work. No significant correlations are
found, independently of the type of best-fit model, thus confirming
the results of \citet{vdb11,vdb12} and his suggestion that the
large observed scatter is probably due to the spread of cluster orbital
parameters, not being correlated with other structural/chemical
features.

Finally, our catalog allows us to
discuss the distribution of the ratio between the core radius and the
effective radius ($r_\mathrm{c}/r_\mathrm{e}$).
%that we will henceforth indicate as $r_\mathrm{c}/r_\mathrm{e}$.
Standard dynamical models of GCs suggest that the value of this parameter
tends to decrease during the cluster long-term evolution driven by two-body
relaxation, until an energy source (e.g. primordial binaries or three-body binaries)
halts the core contraction and this ratio settles to a value determined by
the efficiency of the energy source \citep[see e.g.][]{heggie03}. On
the other hand it has been shown that the presence of exotic
populations, such as stellar mass black holes or an IMBH in the core of GCs, may
prevent the decrease of this ratio and, possibly, cause its increase
\citep[e.g.][]{merritt04,baum05,heggie07, mackey08, trenti10}. For this reason,
such a ratio has been used for preliminary selection of GCs that might harbor an IMBH,
e.g. in deep radio imaging studies \citep{strader12}.

\begin{figure}
%\plotone{kappa.ps}
\includegraphics[width= 8.3 truecm]{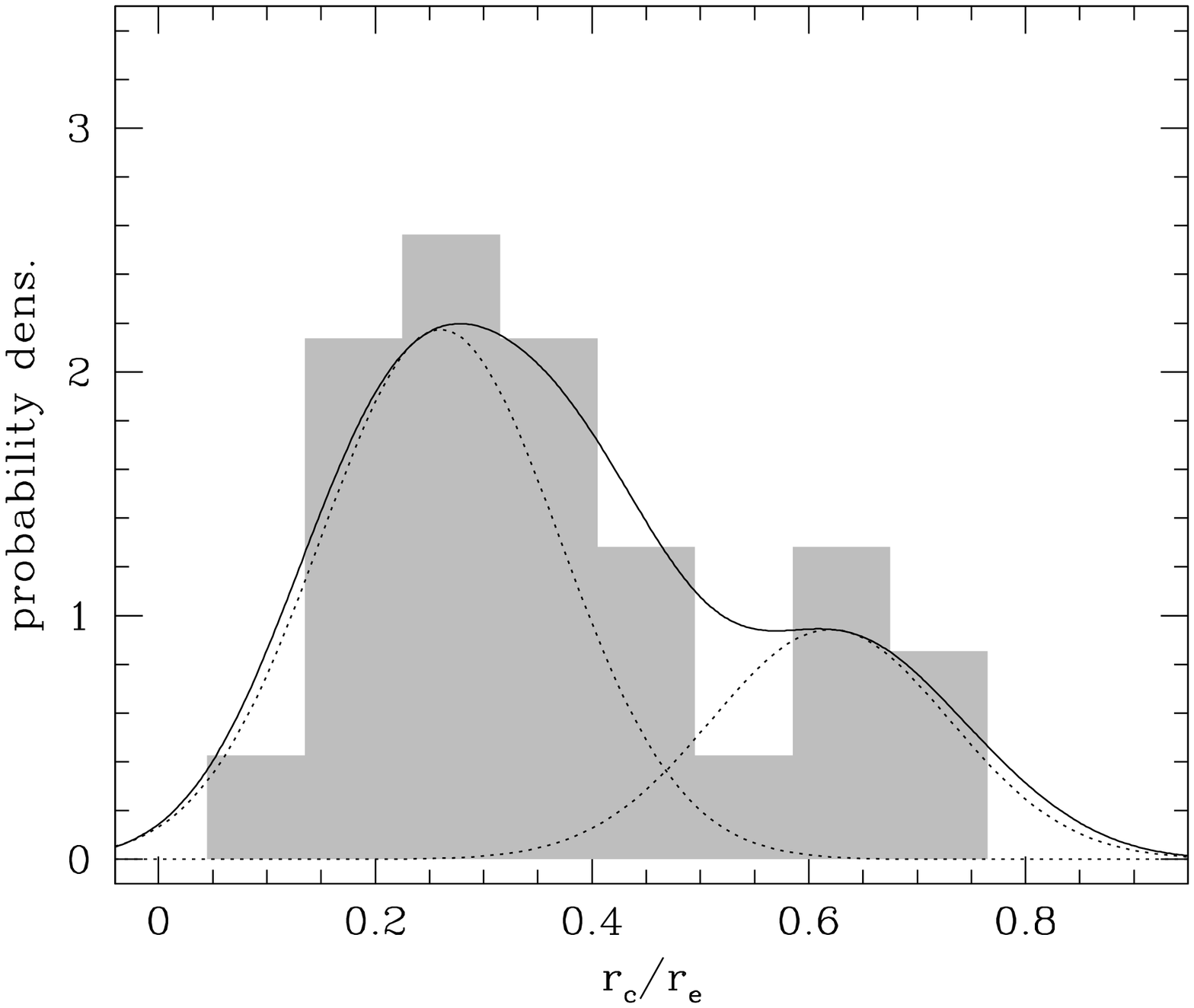}
\caption{
Histogram of the probability density distribution of the best-fit
$r_\mathrm{c}/r_\mathrm{e}$ values for the GCs in our sample.
The solid line corresponds to the underlying probability density function
as deduced from Kernel Density Estimation.
The dotted lines represent two Gaussians with the same standard
deviation ($\sigma = 1.1$) that well fit, by eye, the two observed maxima.
\label{rcsurcKDE}}
\end{figure}

The histogram in Fig.~\ref{rcsurcKDE} shows the distribution of 
$r_\mathrm{c}/r_\mathrm{e}$ as turns out from our GC sample. Indeed, the distribution
appears to be bimodal or at least significantly tailed toward high values.
In order to provide a quantitative statistical support to this appearance we
reconstruct the distribution of $r_\mathrm{c}/r_\mathrm{e}$ using the
Kernel Density Estimation \citep{KDEref3, KDEref2, KDEref1}.  This
technique is essentially a generalized histogram, that allows to
non-parametrically recover the underlying distribution of a
variable based on a sample of $n$ points by adding together $n$ \emph{bump}
functions (kernels) centered on each point.  Fig.~\ref{rcsurcKDE}
shows the probability-density distribution thus obtained: a
qualitative indication of bimodality emerges.  The probability density
is well-reproduced by the superposition of two Gaussian distributions
with the same standard deviation, suggesting that two different
populations exist (a more compact one with $r_\mathrm{c}/r_\mathrm{e}$ peaked
around $0.26$, and a less compact group peaked around $0.62$) and are barely resolved due to observational errors
on both parameters. We run a Shapiro-Wilk normality test \citep{ShaW2,
 ShaW1} obtaining that, under the null-hypothesis of normality, the
$p$-value for our data is relatively low ($p = 0.099$).  While this is
not, by itself, a strong indication that the null-hypothesis of the
data coming from a single normal distribution is to be rejected, we
also note that the skewness and kurtosis of the distribution, as
estimated from the sample, are $0.57$ and $-0.86$ respectively (as
opposed to an expected value of $0$ in the normal case). Then, despite
the not very large number of clusters in our sample, we can conclude
against normality, arguing that the underlying distribution is bimodal
or at least heavy tailed.

%\clearpage

\begin{figure}
%\plotone{kappa.ps}
\includegraphics[width= 8.6 truecm]{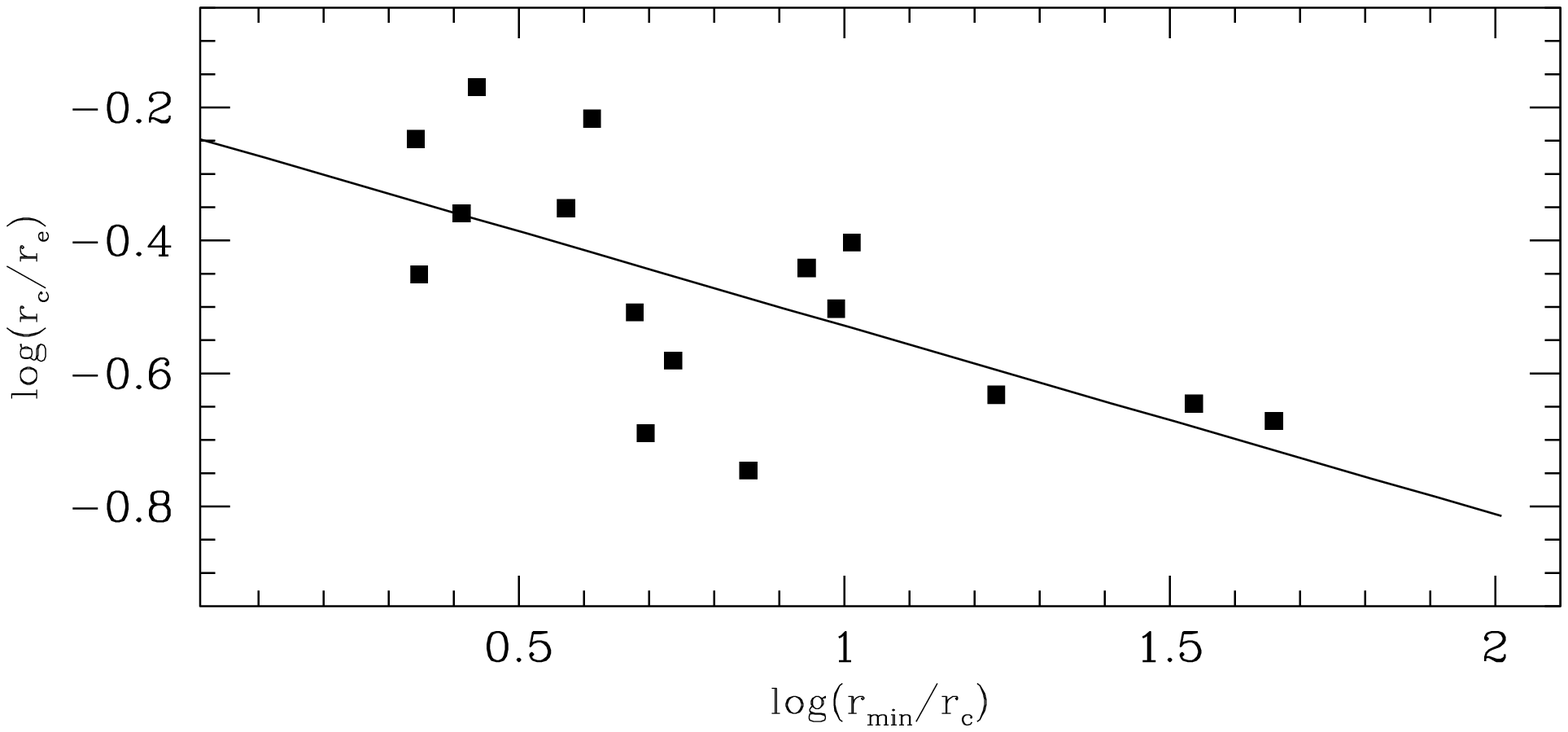}
%{rce_age.ps}
\caption{
Core to effective radii ratio as a function of
the \citet{ferraro12} ``dynamical clock'' parameter, plotted
for the 14 clusters in common with this study, plus two additional
cases, namely NGC 5466 (Beccari et al., in prep.) and NGC 5824 (Sanna et al., in prep.).
Reported is the linear least square fit (solid line).
\label{rce_age}}
\end{figure}

It is interesting to notice that the main peak at $\sim 0.26$ coincides with the value
assumed by $r_\mathrm{c}/r_\mathrm{e}$ during a large fraction of cluster evolution in
the $N$-body simulations of \citet{trenti10}.
As for the group characterized by larger values
of $r_\mathrm{c}/r_\mathrm{e}$, these might be dynamically younger clusters, with values
of $r_\mathrm{c}/r_\mathrm{e}$ corresponding to those imprinted by formation and early
evolution processes. In order to probe this, in Fig. \ref{rce_age} we plot
$r_\mathrm{c}/r_\mathrm{e}$ as a function of the ``dynamical clock''
parameter\footnote{It corresponds to the position
of the minimum ($r_\mathrm{min}$) in the observed blue straggler star
radial distributions, in units of $r_\mathrm{c}$. This radius has been suggested
to progressively move outward (because of dynamical friction) as the cluster
becomes dynamically older. Hence, large values of $r_\mathrm{min}/r_\mathrm{c}$ correspond
to large dynamical ages.},
which has been recently proposed as an empirical
indicator of the cluster dynamical age \citep{ferraro12}.
The well defined trend between $r_\mathrm{c}/r_\mathrm{e}$ and this dynamical-age
indicator does indeed seem to lend support to this interpretation.
As discussed, for example, in \citet{trenti10}, large values
of $r_\mathrm{c}/r_\mathrm{e}$ for dynamically old clusters might require the presence
of an IMBH as an energy source in the cluster core. Although the
characterization of the dynamical age of a cluster is not simple and
much caution is needed in the interpretation of these trends, our
analysis and in particular the absence of any clusters with
large dynamical age (large $r_\mathrm{min}/r_\mathrm{c}$) and large values
of $r_\mathrm{c}/r_\mathrm{e}$ suggests that no IMBH is required
in any clusters of our sample.

%\clearpage

\acknowledgments

This research is part of the Cosmic-Lab project
(\url{http://www.cosmic-lab.eu}) funded by the European Research
Council under contract ERC-2010-AdG-267675.
MP acknowledges the support provided by the National Research Foundation of Korea to
the Center for Galaxy Evolution Research, and also by the KASI-Yonsei Joint Research
Program for the Frontiers of Astronomy and Space Science and the DRC program of
Korea Research Council of Fundamental Science and Technology (FY 2012).
GB acknowledges the European Community’s Seventh Framework Programme under grant
agreement no. 229517. Finally, the authors want to thank the anonymous referee for 
useful comments and suggestions.

%\clearpage
\appendix

\section{Some details on the parametric models}
\label{app1}
The spherical and single-mass \citet{king66} model in the isotropic form adopts a 
stars energy distribution function (DF), $f_\mathrm{K}(E)$, of the form
\begin{equation}
f_\mathrm{K}(E)\propto\left\{
\begin{array}{ll}
\exp(-E/\sigma^2)-1,& \textrm{if $E<0$,}\\
 0, & \textrm{if $E\geq 0$,}
\end{array}\right.
\label{kingdf} 
\end{equation}
with $\sigma$ being a velocity scale parameter and $E$ the star
total energy. This DF abruptly cuts off at energy $E=0$.
In the isotropic \citet{wilson75} model the DF is slightly changed as:
\begin{equation}
%f(E)\propto\left\{
f_\mathrm{W}(E)\propto\left\{
\begin{array}{ll}
\exp(-E/\sigma^2)-1+E/\sigma^2,& \textrm{if $E<0$,}\\
 0, & \textrm{if $E\geq 0$.}
\end{array}\right.
\label{wilsondf}
\end{equation}
It eliminates the discontinuity of the first derivative that $f_\mathrm{K}(E)$
exhibits at $E=0$ and it decreases more slowly than $f_\mathrm{K}(E)$
for increasing energy, thus going to zero more smoothly.
In practice, the DF of Eq. (\ref{wilsondf})
produces a more extended envelope and a larger effective radius.

More details and comparison between these distribution functions can be found
in Sect. 4.1 of \citetalias{mclaugh}. Here, we just want to remind that in the
numerical solution of the Poisson integration needed to generate self-consistent
parametric models of a given DF, it is quite useful to express the volume density
as a function of the gravitational potential $\Psi(r)$:
\begin{equation}
\rho(r)\propto\int f(\Psi(r)+v^2/2) v^2\ud v 
\end{equation}
(under the assumption
of isotropic velocity $\vv$ distribution and with $v=|\vv|$).
%as a function of the gravitational potential $\Psi(r)$.  
Indeed, for the Wilson model the DF in Eq.(\ref{wilsondf}) leads to
\begin{equation}
\rho(W)=\rho_1\left[ \e^W\erf(\sqrt{W})-(4W/\pi)^{1/2}\left(1+\frac{2W}{3}\right)\right]
+\,4\rho_1 \frac{W^2}{15},
\end{equation}
% \[
% \rho(W)=\rho_1\left[ \e^W\erf(\sqrt{W})-(4W/\pi)^{1/2}\left(1+\frac{2W}{3}\right)\right]
% \]
% \begin{equation}
% \ \ \ +\,4\rho_1 \frac{W^2}{15},
% \end{equation}
where the first term is the well known formula giving the volume density
for the King model \citep[see, e.g.,][]{BT},
$W=W(r)\equiv -\Psi(r)/\sigma^2$ is the dimensionless potential,
$\erf(x)=(2/\sqrt{\pi})\times\int_0^x e^{-t^2}\ud t$
is the error function and $\rho_1$ is a normalisation factor.
The various scale parameters satisfy, in both types of model, the relation
$9\sigma^2=4\pi Gr_0^2\rho(W_0)$ with $W_0\equiv W(0)$. 

\begin{figure}
\center
\includegraphics[width= 12.5 truecm]{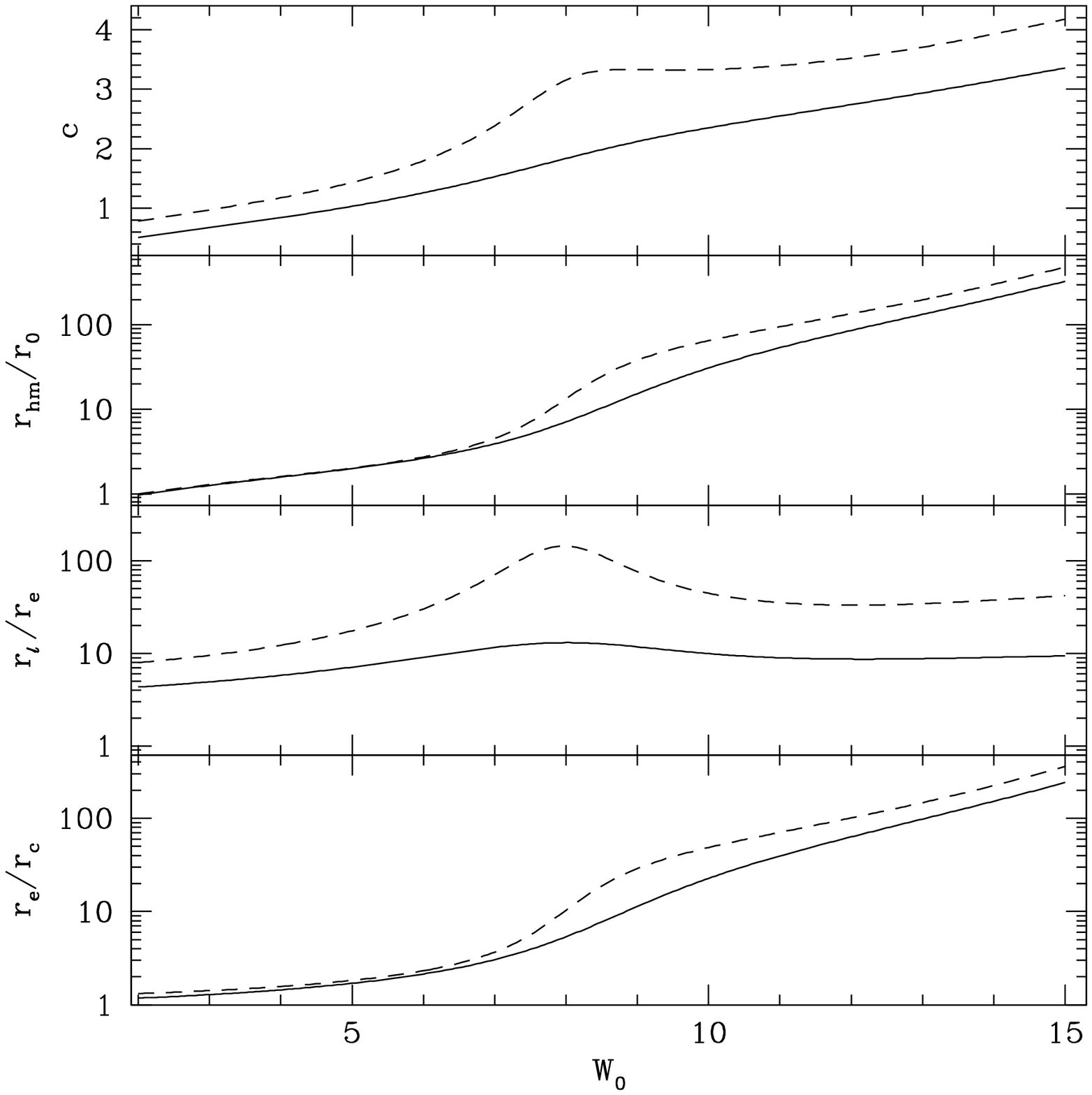}
\caption{Behaviors of various structural parameter as a function of the shape
parameter $W_0$ for both types of model (King: solid line, Wilson: dashed line).}
\label{apprel}
\end{figure}

In Fig.~\ref{apprel} some relevant relations among various parameters
are reported for both models, as a function of the dimensionless central
potential. These relations confirm that Wilson model
yields larger envelopes. In fact, larger values of $r_\l/r_\mathrm{e}$  and of $c$
are found at any given $W_0$ in the Wilson model with respect to the King one.
Moreover, also $r_\mathrm{e}/r_\mathrm{c}$ and $r_\mathrm{hm}/r_0$
are systematically larger in the Wilson model. Notice, finally, that the ratio $r_\l/r_\mathrm{e}$ 
is a limited quantity, 
i.e. $r_\l/r_\mathrm{e}\lesssim 13$ for the King and $r_\l/r_\mathrm{e}\lesssim 145$
for the Wilson model.

%\clearpage

% \clearpage
% \begin{figure}
% %\plotone{kappa.ps}
% \includegraphics[width= 14.5 truecm]{rle_rg.ps}
% \caption{Left panel: cluster limiting to effective radii ratio as a function
% of the galactocentric distance. Right panel: the histogram of the distribution
% of this ratio. The dashed line (left panel) at $log(r_\l/r_\mathrm{e})=1.3$ separates
% two populations having either large or small envelope size.
% The parameters are taken from the best-fit type of model (i.e. that giving the
% lowest $\chi^2_\nu$ for each cluster).
% \label{rle_rg}}
% \end{figure}

% \clearpage
% \begin{figure}
% %\plotone{kappa.ps}
% \includegraphics[width= 14.5 truecm]{fig_thrmin.ps}
% \caption{
% The same quantity of Fig.~\ref{rce_age} as a function
% of the half-mass relaxation time (in units of Hubble time) as
% evaluated in \citet{ferraro12}. \textbf{forse sarebbe piu' giusto
% usare il trh valutato ``indipendentemente'' con la formula diretta}
% \label{rce_age2}}
% \end{figure}

%\label{lastpage}
%\onecolumn
%\bsp

%%%%%%%%%%%%%%%%%%%%%%INIZIO TABELLONE %%%%%%%%%%%%%%%%%%%%%%%%%%%%%%%%%
\clearpage
\LongTables
%\begin{deluxetable*}{@{}lcccccccccc@{}}
\begin{deluxetable*}{lcccccccccc}
%\begin{deluxetable}{@{}lcccccccccc@{}}
%\tabletypesize{\scriptsize}
%\rotate
\tablecaption{Best-fit structural parameters \label{bigtab}}
\tablewidth{0pt}
\tablehead{
\colhead{NGC no.} & \colhead{model} &\colhead{$\chi^{2}_\nu$} & \colhead{$W_0$}  & \colhead{$c$}  &\colhead{$r_0$} &\colhead{$r_\mathrm{hm}$} &\colhead{$r_\l[\arcmin]$} & \colhead{$r_\mathrm{c}$} & \colhead{$r_\mathrm{e}$} & \colhead{$N_\mathrm{BG}$}
}
%\hline
\startdata
104   (47 Tuc)                                               & K               &$1.1$         & $8.10^{+0.05}_{-0}$   & $1.86^{+0.02}_{-0}$          & $29.0^{+0}_{-0.2}$      & $213^{+7}_{-0}$         & $35^{+1}_{-0}$             & $28.1^{+0}_{-0.2}$     & $156^{+4}_{-0}$           & $0$ \\
%104   (47 Tuc)                                               & K               &$1.3$         & $8.45^{+0.05}_{-0}$   & $1.97^{+0.02}_{-0}$          & $23.1^{+0}_{-0.4}$      & $221^{+6}_{-0}$         & $35.7^{+0.6}_{-0}$         & $22.4^{+0}_{-0.4}$     & $162^{+3}_{-0}$           & $0$ \\                                             
\\
\nod                                                         & W               &$1.5$         & $7.90^{+0.05}_{-0.2}$ & $3.10^{+0.03}_{-0.1}$        & $31.5^{+0.7}_{-0.2}$    & $350^{+20}_{-60}$       & $660^{+40}_{-100}$         & $29.7^{+0.5}_{-0.2}$   & $260^{+20}_{-40}$         & $0$ \\
%\nod                                                         & W               &$1.1$         & $8.15\pm 0.05$        & $3.23\pm 0.02$               & $25.1\pm 0.2$           & $390^{+30}_{-20}$       & $710^{+20}_{-30}$          & $23.9\pm 0.2$          & $290\pm 20$               & $0$ \\
\cline{2-11}\\
                                                              
288                                                          & K               &$1.7$         & $5.80^{+0.05}_{-0}$   & $1.21^{+0.01}_{-0}$          & $80^{+0}_{-20}$         & $190^{+0}_{-50}$        & $21^{+0}_{-6}$             & $70^{+0}_{-20}$        & $140^{+0}_{-40}$          & $2$ \\
\\
\nod                                                         & W               &$0.15$        & $3.65^{+0.05}_{-0.2}$ & $1.10^{+0.01}_{-0.05}$       & $123^{+9}_{-2}$         & $178.0^{+2}_{-0.3}$     & $25.8^{+0.3}_{-1}$         & $91.7^{+4}_{-0.7}$     & $135.5^{+2}_{-0.3}$       & $2$ \\
\cline{2-11}\\

%1851                                                         & K               &$0.21$        & $8.70\pm 0.05$        & $2.04\pm 0.01$               & $5.03^{+0.03}_{-0.05}$  & $59\pm 2$               & $9.2\pm 0.2$               & $4.91^{+0.03}_{-0.05}$ & $43\pm 1$                 & $0$ \\
1851                                                         & K               &$0.88$        & $8.4\pm 0.2$          & $1.95\pm 0.04$               & $5.6\pm 0.1$            & $51\pm 5$               & $8.3^{+0.7}_{-0.6}$        & $5.4\pm 0.1$           & $38^{+4}_{-3}$            & $0$ \\
\\
%\nod                                                         & W               &$0.43$        & $8.90^{+0.05}_{-0.1}$ & $3.3297^{+0.0005}_{-0}$      & $5.33^{+0.09}_{-0.04}$  & $180^{+5}_{-10}$        & $190^{+3}_{-1}$            & $5.13^{+0.08}_{-0.04}$ & $131^{+4}_{-8}$           & $0$ \\
\nod                                                         & W               &$1.4$         & $8.7\pm 0.2$          & $3.327^{+0.003}_{-0.02}$     & $5.8^{+0.2}_{-0.1}$     & $170^{+20}_{-30}$       & $204.13^{+0.03}_{-3}$      & $5.5\pm 0.1$           & $120^{+10}_{-20}$         & $0$ \\
\cline{2-11}\\
                                                              
1904 (M79)                                                   & K               &$0.86$        & $7.75^{+0.05}_{-0.1}$ & $1.76^{+0.02}_{-0.03}$       & $9.8^{+0.6}_{-0.3}$     & $56.657^{+0}_{-0.005}$  & $9.32^{+0.04}_{-0.09}$     & $9.4^{+0.6}_{-0.3}$    & $41.68^{+0.08}_{-0.03}$   & $3$ \\
\\
\nod                                                         & W               &$1.8$         & $6.7^{+0.2}_{-0.1}$   & $2.14^{+0.10}_{-0.06}$       & $12.1^{+0.5}_{-0.7}$    & $42.3^{+1}_{-0.5}$      & $28^{+5}_{-3}$             & $11.0^{+0.4}_{-0.6}$   & $31.8^{+0.9}_{-0.4}$      & $3$ \\
\cline{2-11}\\
                                                              
2419                                                         & K               &$2.1$         & $6.95^{+0.05}_{-0}$   & $1.51^{+0.02}_{-0}$          & $17^{+0}_{-7}$          & $60^{+0}_{-20}$         & $9^{+0}_{-3}$              & $16^{+0}_{-6}$         & $50^{+0}_{-20}$           & $2$ \\
\\
\nod                                                         & W               &$0.090$       & $5.8\pm 0.2$          & $1.73^{+0.07}_{-0.06}$       & $22.0^{+0.7}_{-0.6}$    & $55\pm 1$               & $20^{+3}_{-2}$             & $19.3\pm 0.4$          & $41.6^{+1.0}_{-0.8}$      & $2$ \\
\cline{2-11}\\
                                                              
5024 (M53)                                                   & K               &$5.7$         & $7.55^{+0.05}_{-0}$   & $1.70^{+0.02}_{-0}$          & $23.3^{+0}_{-0.6}$      & $118.9^{+0.2}_{-0}$     & $19.3^{+0.2}_{-0}$         & $22.4^{+0}_{-0.6}$     & $87.8^{+0.3}_{-0}$        & $3$ \\
\\
\nod                                                         & W               &$0.57$        & $6.60\pm 0.05$        & $2.11\pm 0.03$               & $27.3\pm 0.4$           & $93.1^{+0.9}_{-0.8}$    & $59^{+4}_{-3}$             & $24.8\pm 0.3$          & $70.0\pm 0.7$             & $3$ \\
\cline{2-11}\\

%5139 ($\Omega$ Cen) IMBH\\
%\\
%\cline{2-11}
%\\
                                                              
5272 (M3)                                                    & K               &$2.3$         & $8.05\pm 0.05$        & $1.85\pm 0.02$               & $23.5^{+0.7}_{-0.6}$    & $166.7^{+1}_{-0.8}$     & $27.6\pm 0.2$              & $22.7^{+0.7}_{-0.6}$   & $122.1^{+1}_{-0.8}$       & $1$ \\
\\
\nod                                                         & W               &$0.13$        & $6.8\pm 0.1$          & $2.28\pm 0.07$               & $28.6^{+0.8}_{-0.7}$    & $112^{+6}_{-3}$         & $90\pm 10$                 & $26.2\pm 0.6$          & $85\pm 3$                 & $1$ \\
\cline{2-11}\\
                                                              
5466                                                         & K               &$3.0$         & $6.2\pm 0.1$          & $1.31\pm 0.03$               & $78\pm 3$               & $214\pm 2$              & $26.3^{+0.5}_{-0.4}$       & $72\pm 3$              & $160^{+2}_{-1}$           & $3$ \\
\\
\nod                                                         & W               &$0.95$        & $5.0^{+0.2}_{-0}$     & $1.42^{+0.04}_{-0}$          & $100^{+0}_{-70}$        & $200^{+0}_{-100}$       & $40^{+0}_{-30}$            & $80^{+0}_{-50}$        & $150^{+0}_{-90}$          & $3$ \\
\cline{2-11}\\

5824                                                         & K               &$6.2$         & $8.95^{+0.05}_{-0}$   & $2.11^{+0.01}_{-0}$          & $4.1^{+0}_{-0.1}$       & $58.6^{+0.8}_{-0}$      & $8.793^{+0}_{-0.003}$      & $4.0^{+0}_{-0.1}$      & $42.7^{+0.3}_{-0}$        & $3$ \\
\\
\nod                                                         & W               &$0.21$        & $7.4\pm 0.1$          & $2.71^{+0.09}_{-0.08}$       & $4.8\pm 0.2$            & $29\pm 2$               & $40^{+7}_{-5}$             & $4.4^{+0.2}_{-0.1}$    & $22^{+2}_{-1.0}$          & $3$ \\
\cline{2-11}\\

5904 (M5)                                                    & K               &$1.5$         & $7.45^{+0.05}_{-0.1}$ & $1.66\pm 0.02$               & $29^{+0}_{-8}$          & $140^{+0}_{-40}$        & $23^{+0}_{-7}$             & $28^{+0}_{-7}$         & $100^{+0}_{-30}$          & $2$ \\ %rifatto da B 7/05/13
%5904 (M5)                                                    & K               &$2.2$         & $7.65\pm 0.05$        & $1.73\pm 0.02$               & $25.9^{+0.7}_{-0.8}$    & $140.9^{+0.2}_{-2}$     & $23.0^{+0.1}_{-0.2}$       & $24.9^{+0.6}_{-0.7}$   & $103.6^{+0.1}_{-0.5}$     & $2$ \\
\\
\nod                                                         & W               &$1.3$         & $6.55\pm 0.05$        & $2.08\pm 0.03$               & $32.1\pm 0.5$           & $107^{+1}_{-2}$         & $65^{+4}_{-3}$             & $29.1\pm 0.4$          & $80.5\pm 0.8$             & $2$ \\ %rifatto da B 7/05/13
%\nod                                                         & W               &$1.3$         & $6.70^{+0.05}_{-0.1}$ & $2.17^{+0.03}_{-0.06}$       & $33.4^{+1}_{-0.5}$      & $120\pm 1$              & $83^{+5}_{-9}$             & $30.4^{+1}_{-0.4}$     & $90\pm 1$                 & $2$ \\
\cline{2-11}\\

%6093 (M80) & tolto\\
%\cline{2-11}\\

6121 (M4)                                                    & K               &$0.41$        & $7.5\pm 0.2$          & $1.68\pm 0.06$               & $67\pm 3$               & $330^{+30}_{-20}$       & $53^{+6}_{-5}$             & $64^{+3}_{-2}$         & $240\pm 20$               & $2$ \\
\\
\nod                                                         & W               &$0.68$        & $7.8\pm 0.2$          & $3.0\pm 0.2$                 & $69\pm 3$               & $700\pm 200$            & $1200^{+500}_{-400}$       & $65\pm 2$              & $500^{+200}_{-100}$       & $2$ \\
\cline{2-11}\\

6205 (M13)                                                   & K               &$0.49$        & $6.2^{+0.2}_{-0.1}$   & $1.32^{+0.04}_{-0.03}$       & $53\pm 1$               & $149^{+4}_{-2}$         & $18.5^{+1}_{-0.8}$         & $49.5^{+0.7}_{-1}$     & $111^{+3}_{-2}$           & $2$ \\%agg. 29/04
%6205 (M13)                                                   & K               &$1.2$         & $6.9^{+0.2}_{-0.1}$   & $1.50^{+0.04}_{-0.03}$       & $36\pm 2$               & $129.0^{+0.6}_{-0.2}$   & $18.7^{+0.6}_{-0.3}$       & $34\pm 2$              & $96.0^{+0.5}_{-0.1}$      & $4$ \\
\\
\nod                                                         & W               &$0.69$        & $6.0\pm 0.2$          & $1.77^{+0.07}_{-0.08}$       & $57^{+2}_{-1}$          & $148^{+5}_{-6}$         & $57\pm 9$                  & $50.5^{+0.9}_{-0.6}$   & $112\pm 4$                & $2$ \\%agg. 29/04
%\nod                                                         & W               &$0.64$        & $6.0\pm 0.2$          & $1.80^{+0.07}_{-0.09}$       & $44^{+3}_{-2}$          & $115.2^{+1.0}_{-0.8}$   & $46\pm 6$                  & $39\pm 2$              & $87.0^{+0.9}_{-0.7}$      & $4$ \\
\cline{2-11}\\
                                                              
6229\tablenotemark{a}                                        & K               &$8.1$         & $7.40\pm 0.05$        & $1.65\pm 0.02$               & $8.3\pm 0.2$            & $38.6^{+0}_{-0.2}$      & $6.12^{+0.04}_{-0.05}$     & $7.9\pm 0.2$           & $28.50^{+0}_{-0.01}$    & $3$ \\
\\
\nod                                                         & W               &$0.72$        & $6.05\pm 0.05$        & $1.82\pm 0.02$               & $10.9\pm 0.2$           & $29.27^{+0.08}_{-0.04}$ & $12.0^{+0.5}_{-0.4}$       & $9.7\pm 0.1$           & $22.12^{+0.07}_{-0.04}$   & $3$ \\
\cline{2-11}\\

6254 (M10)                                                   & K               &$0.091$       & $6.6\pm 0.1$          & $1.41\pm 0.03$               & $44\pm 2$               & $139.9^{+1}_{-0.1}$     & $19.0^{+0.6}_{-0.5}$       & $41\pm 1$              & $104.7^{+0.4}_{-0.1}$     & $4$ \\
\\
\nod                                                         & W               &$0.48$        & $6.0\pm 0.2$          & $1.80\pm 0.07$               & $50\pm 2$               & $132^{+2}_{-1}$         & $52^{+7}_{-6}$             & $44^{+2}_{-1}$         & $100^{+2}_{-1}$           & $4$ \\
\cline{2-11}\\
                                                              
%6266 (M62)                                                   & K               &$0.55$        & $7.5\pm 0.1$          & $1.66\pm 0.03$               & $17.4\pm 0.8$           & $83.5^{+0.7}_{-1}$      & $13.4\pm 0.3$              & $16.6^{+0.7}_{-0.8}$   & $61.6\pm 0.6$             & $18$ \\
6266 (M62)                                                   & K               &$0.27$        & $7.8\pm 0.2$          & $1.79\pm 0.05$               & $15.9^{+0.7}_{-0.6}$    & $99^{+7}_{-5}$          & $16\pm 1$                  & $15.4\pm 0.6$          & $72^{+5}_{-4}$            & $4$ \\
\\
%\nod                                                         & W               &$1.2$         & $6.5^{+0.8}_{-0.2}$   & $2.08^{+0.5}_{-0.09}$        & $22^{+1}_{-4}$          & $74^{+30}_{-2}$         & $45^{+80}_{-6}$            & $20.0^{+0.8}_{-3}$     & $55^{+20}_{-1}$           & $18$ \\
\nod                                                         & W               &$0.48$        & $8.0\pm 0.2$          & $3.1^{+0.1}_{-0.10}$         & $16.8^{+0.4}_{-0.7}$    & $200^{+70}_{-30}$       & $380^{+100}_{-70}$         & $15.8^{+0.3}_{-0.6}$   & $150^{+50}_{-20}$         & $4$ \\
\cline{2-11}\\

%6388 IMBH \\
%\\
%\cline{2-11}
%\\
%6624 patol.\\
%\\
%\cline{2-11}
%\\
%6715 (M54) IMBH \\
%\\
%\cline{2-11}
%\\

6341 (M92)                                                   & K               &$0.41$        & $7.70^{+0.1}_{-0.05}$ & $1.74^{+0.03}_{-0.02}$       & $15.2^{+0.3}_{-0.5}$    & $85^{+3}_{-1}$          & $13.9^{+0.5}_{-0.2}$       & $14.6^{+0.2}_{-0.5}$   & $62.6^{+2}_{-0.9}$        & $3$ \\
\\
\nod                                                         & W               &$0.54$        & $6.70\pm 0.05$        & $2.17\pm 0.03$               & $20^{+20}_{-0}$         & $67^{+60}_{-1.0}$       & $46^{+50}_{-3}$            & $20^{+20}_{-0}$        & $50.2^{+50}_{-0.8}$       & $3$ \\
\cline{2-11}\\
                                                              
6626 (M28)                                                   & K               &$0.59$        & $8.6\pm 0.2$          & $2.01^{+0.07}_{-0.06}$       & $10.8\pm 0.3$           & $120^{+20}_{-10}$       & $19^{+3}_{-2}$             & $10.5^{+0.2}_{-0.3}$   & $90^{+20}_{-10}$          & $4$ \\%aggiornato 14/05/13
\\
\nod                                                         & W               &$0.94$        & $9.1\pm 0.2$          & $3.329^{+0.001}_{-0.004}$    & $11.0\pm 0.2$           & $410^{+50}_{-60}$       & $390\pm 10$                & $10.6\pm 0.2$          & $300^{+40}_{-50}$         & $4$ \\% aggiornato 14/05/13
\cline{2-11}\\

6809 (M55)                                                   & K               &$0.68$        & $5.0\pm 0.2$          & $1.02^{+0.04}_{-0.05}$       & $113^{+7}_{-5}$         & $216\pm 1$              & $20\pm 1$                  & $99^{+4}_{-3}$         & $162.8^{+0.6}_{-0.1}$     & $3$ \\
\\
\nod                                                         & W               &$1.1$         & $4.3^{+0.3}_{-0.4}$   & $1.24\pm 0.09$               & $128^{+10}_{-8}$        & $213^{+4}_{-2}$         & $37^{+6}_{-5}$             & $101\pm 4$             & $162^{+3}_{-2}$           & $3$ \\
\cline{2-11}\\
                                                              
6864 (M75)\tablenotemark{b}                                  & K               &$0.49$        & $7.85\pm 0.05$        & $1.79\pm 0.02$               & $5.1^{+0.1}_{-2}$       & $30^{+0}_{-10}$         & $5^{+0}_{-2}$              & $4.9^{+0.1}_{-2}$      & $23^{+0}_{-8}$            & $4$ \\
\\
\nod                                                         & W               &$0.80$        & $7.0^{+0.2}_{-0.1}$   & $2.38^{+0.1}_{-0.07}$        & $6.2^{+0.2}_{-0.3}$     & $27.0^{+2}_{-0.8}$      & $25^{+6}_{-3}$             & $5.7^{+0.2}_{-0.3}$    & $20.4^{+1}_{-0.7}$        & $4$ \\
\cline{2-11}\\
                                                              
7089 (M2)\tablenotemark{c}                                   & K               &$0.35$        & $7.15^{+0.2}_{-0.05}$ & $1.57^{+0.04}_{-0.02}$       & $16.2^{+0.4}_{-6}$      & $66.3^{+0.2}_{-20}$     & $10.1^{+0.2}_{-3}$         & $15.4^{+0.4}_{-5}$     & $49.1^{+0.2}_{-20}$       & $4$ \\
\\
\nod                                                         & W               &$0.99$        & $6.5^{+0.1}_{-0.2}$   & $2.02^{+0.06}_{-0.1}$        & $17.5^{+1}_{-0.5}$      & $55\pm 2$               & $31^{+4}_{-5}$             & $15.8^{+0.8}_{-0.4}$   & $41.8^{+0.8}_{-1}$        & $4$ \\
\cline{2-11}\\

%\\
%7099 (M30) patol.\\
%\\

Am 1                                                         & K               &$0.53$        & $7.1^{+0.6}_{-0.5}$   & $1.6^{+0.2}_{-0.1}$          & $10\pm 2$               & $39^{+5}_{-1}$          & $5.8^{+1}_{-0.7}$          & $9^{+2}_{-1}$          & $28.5^{+4}_{-0.5}$        & $3$ \\
\\
\nod                                                         & W               &$0.24$        & $6.5^{+0.7}_{-0.8}$   & $2.0^{+0.5}_{-0.3}$          & $11\pm 2$               & $34^{+10}_{-3}$         & $19^{+30}_{-9}$            & $10^{+2}_{-1}$         & $26^{+8}_{-2}$            & $3$ \\
\cline{2-11}\\
                                                              
Eridanus                                                     & K               &$0.18$        & $6.4^{+1}_{-0.8}$     & $1.4^{+0.3}_{-0.2}$          & $16^{+2}_{-1}$          & $47^{+20}_{-7}$         & $6^{+5}_{-2}$              & $14.8\pm 0.9$          & $35^{+20}_{-5}$           & $0$ \\
\\
\nod                                                         & W               &$0.14$        & $7\pm 1$              & $2.2^{+0.9}_{-0.5}$          & $16\pm 2$               & $60^{+100}_{-20}$       & $40^{+300}_{-30}$          & $14.9^{+1.0}_{-0.9}$   & $40^{+80}_{-10}$          & $0$ \\
\cline{2-11}\\
                                                              
Pal 3                                                        & K               &$0.10$        & $3.7^{+1}_{-0.9}$     & $0.8^{+0.2}_{-0.1}$          & $35^{+9}_{-8}$          & $51.0^{+1}_{-0.2}$      & $3.6^{+1}_{-0.4}$          & $28^{+3}_{-5}$         & $38.7^{+0.9}_{-0.2}$      & $4$ \\
\\
\nod                                                         & W               &$0.061$       & $2.1^{+2}_{-0.1}$     & $0.81^{+0.5}_{-0.02}$        & $50^{+1}_{-20}$         & $50.29^{+2}_{-0.02}$    & $5.33^{+4}_{-0.08}$        & $29.3^{+0.2}_{-5}$     & $38.34^{+1}_{-0.05}$      & $4$ \\
\cline{2-11}\\
                                                              
Pal 4                                                        & K               &$0.29$        & $5.2\pm 0.7$          & $1.1^{+0.2}_{-0.1}$          & $25^{+5}_{-4}$          & $50.5^{+2}_{-0.1}$      & $4.9^{+0.9}_{-0.6}$        & $22^{+4}_{-3}$         & $38.1^{+1}_{-0.2}$        & $3$ \\
\\
\nod                                                         & W               &$0.15$        & $4\pm 1$              & $1.3^{+0.3}_{-0.2}$          & $29^{+9}_{-5}$          & $48.65^{+2}_{-0.010}$   & $9^{+5}_{-2}$              & $23^{+4}_{-3}$         & $37^{+1}_{-0}$            & $3$ \\
\cline{2-11}\\
                                                              
Pal 14                                                       & K               &$0.15$        & $4.3\pm 0.7$          & $0.9\pm 0.1$                 & $48^{+10}_{-8}$         & $80.0^{+2}_{-0.7}$      & $6.4^{+1.0}_{-0.6}$        & $41\pm 5$              & $60.5^{+2}_{-0.5}$        & $4$ \\
\\
\nod                                                         & W               &$0.15$        & $3\pm 1$              & $1.0^{+0.3}_{-0.2}$          & $60^{+20}_{-10}$        & $78.2^{+1}_{-0.4}$      & $10^{+5}_{-2}$             & $42\pm 5$              & $59.4^{+1}_{-0.1}$        & $4$ \\
\cline{2-11}\\
                                                              
Ter 5                                                        & K               &$0.28$        & $7.2\pm 0.2$          & $1.59^{+0.06}_{-0.04}$       & $8.1^{+0.4}_{-0.5}$     & $34^{+2}_{-1}$          & $5.2^{+0.5}_{-0.3}$        & $7.7^{+0.3}_{-0.4}$    & $25.2^{+1}_{-0.7}$        & $10$ \\
\\
\nod                                                         & W               &$0.19$        & $7.0\pm 0.2$          & $2.4^{+0.2}_{-0.1}$          & $8.8^{+0.2}_{-0.3}$     & $40^{+7}_{-3}$          & $39^{+20}_{-8}$            & $8.1^{+0.2}_{-0.3}$    & $30^{+5}_{-2}$            & $10$ \\

\enddata
\tablecomments{
Best-fit structural parameters of the target GCs. For each
cluster, the results for both the King (K) and the Wilson (W)
model fits are given.  $\chi^2_\nu$ is the reduced $\chi^2$ of the
best fits.
%, with the lowest value indicating the model (between the
%King and the Wilson ones) that best reproduces the observations.
$W_0$ is the central dimensionless potential, $c$ the
concentration parameter, $r_0$ the model scale radius,
$r_\mathrm{hm}$ the 3-dimensional half-mass radius, $r_\l$
the limiting radius, while $r_\mathrm{c}$ and $r_\mathrm{e}$ are
the core and the effective radii, respectively. All radii are in
units of arcseconds, with the exception of $r_\l$
which is in arcmin.  The 1-$\sigma$ uncertainties (computed as
discussed in Sect. \ref{bestfit}) are reported for each parameter.
The number ($N_\mathrm{BG}$) of the outermost data points used for
Galactic background determination is given in the last column (a
null value indicates that the data-set was not radially extended
enough to allow such an estimate).
}
\tablenotetext{a}{The point at $r=560\arcsec$ is excluded from the fit.}
\tablenotetext{b}{The point at $r=310\arcsec$ is excluded.}
\tablenotetext{c}{The point at $r=550\arcsec$ is excluded.}
%\end{deluxetable}
\end{deluxetable*}

\end{document}